\documentclass[12pt,preprint]{aastex}
\usepackage{natbib}
\usepackage{pdflscape}
\usepackage{graphicx}

\begin{document}               

\title{The Three Dimensional Properties and Energetics of Radio Jet Driven Outflows\footnote{Some of the data presented herein were obtained at the W.M. Keck Observatory, which is operated as a scientific partnership among the California Institute of Technology, the University of California and the National Aeronautics and Space Administration. The Observatory was made possible by the generous financial support of the W.M. Keck Foundation. }}
\author{Hsin-Yi Shih, Alan Stockton\\ \ \emph{Institute for Astronomy, University of Hawai'i\\ 2680 Woodlawn Dr, Honolulu, HI 96822}}
\email{hsshih@ifa.hawaii.edu, stockton@ifa.hawaii.edu}


\begin{abstract}
Extended emission-line regions (EELRs), found around radio loud sources, are likely outflows driven by one form of powerful AGN feedback mechanism. We seek to constrain  the three-dimensional gas properties and the outflow energetics of the EELRs in this study. We used an integral field unit to observe EELRs around two samples of of radio loud AGNs with similar radio properties but different orientations: a sample of quasars and a sample of radio galaxies. A morphological comparison suggests a scenario where the three-dimensional EELR gas distribution follows rough biconical shapes with wide opening angles. The average extent of the EELRs is $\sim 18.5$ kpc. The estimated average mass of the EELRs, with reasonable assumptions for gas densities, is $\sim 3 \times 10^8$ M$_\odot$, and the average mass outflow rate is $\sim 30$ M$_\odot$/yr. The EELRs around quasars and radio galaxies share similar kinematic properties. Both samples have velocity structures that display a range of complexities and they do not appear to correlate with the jet orientations, and both span a similar range of velocity dispersions. Around 30\% of the detected EELRs show large scale rotational motions, which may have originated from recent mergers involving gas-rich disk galaxies.

\end{abstract}

\section{Introduction and Purpose}

Extended Emission Line Regions (EELRs) are massive clouds of ionized gas surrounding $\sim 1/3$ of steep-spectrum radio loud AGNs, spanning a range of up to tens of kpc \citep[e.g.,][]{1985ApJ...293..120B, 1987ApJ...316..584S}. Most luminous EELRs are found around Fanaroff-Riley type II (FR II) sources, which are quasars and galaxies that have large radio structures often extending far beyond the host galaxies. Such jets that originate in the nucleus must impart a significant amount of energy to the surrounding gas to break out of the host galaxy. It is likely that EELRs often consist of outflowing gas driven by processes associated with the onset of the radio jets \citep{2009ApJ...690..953F}. The mass of the EELRs can reach up to $\sim 10^{10}$ M$_{\odot}$ \citep{2007ApJ...666..794F}. Such massive outflows can play an important role in galaxy evolution by suppressing black hole growth and/or star formation. 

Studying the physical properties of EELRs is critical to understanding their origin and the potential impact they can have on the evolution of their host galaxies. According to the unification scheme, quasars and FR II radio galaxies are the same type of objects viewed along a different line-of-sight.  The goal of this paper is to characterize the 3-D properties of the EELRs through an integral field unit (IFU) study of a matched sample of quasars and radio galaxies selected to have similar radio properties. The observations will also give estimated masses and energies for a large sample of EELRs. 

All the calculations in this paper assume $H_o = 71$, $\Omega_{M} = 0.27$ and $\Omega_{vac} = 0.73$. Any conversion that requires luminosity distances is calculated using Ned Wright's Cosmology Calculator \citep{2006PASP..118.1711W}.

\section{Sample Selection}
\label{sec:sample}

Our goal is to select quasars and radio galaxies with a similar range of radio luminosity, redshift and radio spectral indices. The radio SEDs are assumed to follow the power law $f = \nu^{-\alpha}$, where $f$ is the flux, $\nu$ is the frequency, and $\alpha$ is the spectral index. We searched the NASA Extragalactic Database for radio loud quasars with 365 MHz flux density $> 1$ Jy, redshift between 0.2 and 0.5, $\alpha > 0.5$ (calculated from the 365 MHz and 1400 MHz flux densities), and declination between $-10$ and $60$ degrees. We found 50 quasars that fit these criteria, and matched each quasar to a radio galaxy that is close in redshift, spectral index and radio power. This sample was observed with the SuperNovae Integral Field Spectrograph (SNIFS) \citep{2002SPIE.4836...61A,2004SPIE.5249..146L} mounted on the University of Hawaii 2.2 meter telescope \footnote{SNIFS on the UH 2.2-m telescope is part of the Nearby Supernova Factory project, a scientific collaboration among the Centre de Recherche Astronomique de Lyon, Institut de Physique Nuclaire de Lyon, Laboratoire de Physique Nuclaire et des Hautes Energies, Lawrence Berkeley National Laboratory, Yale University, University of Bonn, Max Planck Institute for Astrophysics, Tsinghua Center for Astrophysics, and the Centre de Physique des Particules de Marseille.}.

Selecting a radio galaxy sample well-matched to the quasar sample is complicated by the fact that the powerful radio galaxy population is more inhomogeneous than the quasar populations. While the simple unification scheme states that FR II sources harbor quasars hidden by dust tori, there is a fraction of FR II radio galaxies that belong to a group of low-excitation radio galaxies (LERGs), which show no evidence of containing obscured quasars \citep{1994ASPC...54..201L, 2006MNRAS.370.1893H}. LERGs are characterized by having low [\ion{O}{3}] emission relative to [\ion{O}{2}] emission and/or the continuum, and some show no detectable emission at all. LERGs likely reside in dense environments where the radio activity is driven by radiatively inefficient accretion from the IGM. The RGs that do contain hidden quasars are high-excitation radio galaxies (HERGs). \citet{2006MNRAS.370.1893H} showed that LERGs at the same radio luminosity as HERGs have accretion luminosities that are $\sim 10$ times smaller than those of the HERGs. Since the occurrence of EELRs depends on the availability of gas around the central region, it is unlikely that the LERGs, with much lower accretion rates, will form massive outflows. Also, since LERGs have weaker accretion disks that produce lower ionizing continua, they are unable to ionize the amount of gas found in typical EELRs. 

The lack of available optical spectra for many of our potential RGs means that LERGs could not be completely selected out before observation. Due to telescope time constraints, we could not observe a significantly larger RG sample to compensate for the inhomogeneity of the RG population. After we retroactively remove the LERGs, the RG and quasar samples are not as well matched in radio power and redshift. Fortunately, we have supplemental longslit observations of 37 FR II radio galaxies from a related project. This sample was observed with the low resolution imaging spectrograph (LRIS) on the Keck telescope \citep{1995PASP..107..375O}. These galaxies were selected to have radio power $> 0.5$ Jy at 365 MHz, redshift between 0.15 and 0.4, and spectral index $> 0.7$. These have lower redshift and higher spectral indices than the sample selected for this project. However, the selection criteria are sufficiently close to ours that we can merge and re-match the two samples (LRIS and SNIFS) where appropriate to check for biases.

\section{Observations}
The SNIFS instrument comprises two channels, blue and red. The blue channel covers $3500 - 5700$ \AA, the red channel covers $5300 - 10500$ \AA. The spectral resolution is R$\sim 1200$. The spatial coverage is $6\arcsec \times 6\arcsec$ with a spatial resolution of $0\farcs4 \times 0\farcs4$. Each object had an exposure time of at least 1200 s on clear nights; the seeing ranged from $0.6\arcsec$ to $1.5\arcsec$. The radio properties and redshift for all observed targets are listed in Table \ref{wholesample}. Given the weather constraints, the final observed sample consists of 39 quasars and 41 radio galaxies. 

The supplemental RG sample was observed with Keck LRIS longslit, using the 600/7500 grating centered at 6800~\AA. We used slits that are either 8.7\arcsec or 1.5\arcsec wide. The integration time on each object was 300 seconds. The sample is listed in Table \ref{lrisobs} along with their physical properties and observation parameters. Since these observations have higher signal to noise than the SNIFS observations, they include some faint EELR detections that would not have been detected in the SNIFS observations. For consistency, such EELRs are not considered as detections.

\begin{deluxetable}{lcccccc}
\tablecaption{Whole Sample}
\tabletypesize{\scriptsize}
\tablecolumns{7}
\tablehead{\colhead{Object Name} & \colhead{Object Type} & \colhead{Redshift}  & \colhead{F(1400 MHz)}  & \colhead{F(365 MHz)} & \colhead{Spectral Index} & \colhead{EELR Detected?}\\
\colhead{} & \colhead{} & \colhead{} & \colhead{Jy} & \colhead{Jy} & \colhead{} & \colhead{}}
\startdata
3C047 * & QSO & 0.43 &  3.75 & 13.02 & 0.93 & yes\\
4C+41.04 & QSO & 0.50 &  1.11 &  2.38 & 0.57 & yes\\
4C+25.40 * & QSO & 0.27 &  0.49 &  1.15 & 0.64 & yes\\
4C+25.01 * & QSO & 0.28 &  0.71 &  1.51 & 0.56 & yes\\
4C+15.09 & QSO & 0.49 &  0.83 &  2.32 & 0.76 & yes\\
4C+02.23 * & QSO & 0.40 &  1.79 &  4.70 & 0.72 & yes\\
4C+09.35 * & QSO & 0.30 &  0.42 &  1.32 & 0.85 & yes\\
4C+22.25 * & QSO & 0.42 &  1.10 &  3.11 & 0.77 & yes\\
4C+31.38 & QSO & 0.42 &  2.78 &  7.20 & 0.66 & yes\\
TXS1220+373 & QSO & 0.49 &  0.46 &  1.44 & 0.84 & yes\\
4C+11.50 * & QSO & 0.44 &  0.60 &  2.04 & 0.91 & yes\\
3C246 * & QSO & 0.35 &  2.16 &  5.97 & 0.76 & yes\\
3C277.1 * & QSO & 0.32 &  2.45 &  7.44 & 0.78 & yes\\
3C093 * & QSO & 0.36 &  2.80 &  9.13 & 0.88 & no\\
3C215 * & QSO & 0.41 &  1.60 &  5.06 & 0.86 & no\\
3C240 & QSO & 0.47 &  1.30 &  3.48 & 0.73 & no\\
4C+21.26 * & QSO & 0.30 &  0.87 &  1.85 & 0.56 & yes\\
4C+22.44 & QSO & 0.46 &  0.49 &  2.02 & 1.05 & no\\
4C+27.48 & QSO & 0.37 &  0.25 &  1.28 & 1.22 & no\\
4C+41.18 * & QSO & 0.41 &  0.83 &  2.30 & 0.75 & no\\
4C+49.25 & QSO & 0.21 &  1.19 &  2.48 & 0.55 & no\\
TXS1128+455 * & QSO & 0.40 &  1.97 &  5.05 & 0.70 & no\\
4C+11.06 * & QSO & 0.23 &  0.62 &  2.00 & 0.87 & no\\
4C+29.02 * & QSO & 0.36 &  0.79 &  2.05 & 0.71 & yes\\
4C+31.06 * & QSO & 0.37 &  1.03 &  2.27 & 0.59 & yes\\
4C+10.30 * & QSO & 0.42 &  0.59 &  2.23 & 0.99 & no\\
TXS1608+113 & QSO & 0.46 &  0.31 &  1.08 & 0.92 & no\\
4C+27.38 * & QSO & 0.37 &  0.90 &  2.13 & 0.64 & yes\\
TXS1745+163 * & QSO & 0.39 &  0.34 &  1.29 & 0.98 & yes\\
TXS1951+498 * & QSO & 0.42 &  0.41 &  1.46 & 0.95 & no\\
4C+08.64 & QSO & 0.48 &  1.69 &  4.89 & 0.72 & yes\\
4C+11.72 * & QSO & 0.33 &  1.60 &  4.39 & 0.75 & no\\
4C+09.72 & QSO & 0.43 &  0.67 &  1.63 & 0.67 & no\\
4C+03.59 * & QSO & 0.27 &  1.50 &  4.20 & 0.76 & no\\
TXS0404+065 * & QSO & 0.35 &  0.36 &  1.05 & 0.80 & no\\
4C-00.43 & QSO & 0.42 &  1.05 &  2.63 & 0.68 & no\\
PG1103-06 & QSO & 0.42 &  1.05 &  2.63 & 0.68 & no\\
3C323.1 * & QSO & 0.26 &  2.48 &  5.04 & 0.53 & yes\\
PKS1509+022 & QSO & 0.22 &  0.99 &  2.36 & 0.65 & no\\
3C099 * & RG & 0.43 &  1.60 &  5.34 & 0.90 & yes\\
3C262 & RG & 0.44 &  2.62 &  6.66 & 0.70 & yes\\
3C299 * & RG & 0.37 &  2.70 &  8.05 & 0.81 & yes\\
3C456 * & RG & 0.23 &  2.50 &  7.99 & 0.85 & yes\\
3C459 & RG & 0.22 &  4.25 & 17.21 & 1.07 & yes\\
4C+09.44 * & RG & 0.25 &  1.59 &  3.70 & 0.63 & yes\\
4C+10.71 * & RG & 0.25 &  1.21 &  2.86 & 0.64 & yes\\
4C+20.27 * & RG & 0.47 &  1.45 &  4.68 & 0.87 & yes\\
4C+22.45 & RG & 0.25 &  1.48 &  3.49 & 0.64 & yes\\
PKS0230-027 * & RG & 0.24 &  0.47 &  1.01 & 0.57 & yes\\
3C073 & RG & 0.20 &  1.87 &  5.80 & 0.84 & no\\
3C166 * & RG & 0.24 &  2.60 &  7.55 & 0.84 & no\\
3C244.1 * & RG & 0.43 &  3.78 & 13.89 & 0.96 & yes\\
3C284 & RG & 0.24 &  0.00 &  7.07 & 1.00 & yes\\
3C290 & RG & 0.24 &  0.70 &  2.17 & 0.91 & no\\
3C438 & RG & 0.29 &  6.80 & 26.40 & 0.99 & no\\
4C+00.40 & RG & 0.21 &  0.91 &  2.59 & 0.78 & no\\
4C+00.46 * & RG & 0.42 &  1.65 &  4.53 & 0.75 & no\\
4C+06.56 & RG & 0.34 &  0.31 &  1.14 & 0.98 & no\\
4C+10.10 & RG & 0.22 &  1.63 &  4.17 & 0.70 & no\\
4C+11.12 & RG & 0.45 &  0.42 &  1.17 & 0.76 & no\\
4C+14.11 & RG & 0.21 &  2.10 &  4.67 & 0.59 & no\\
4C+30.14 & RG & 0.32 &  0.47 &  1.50 & 0.86 & no\\
4C+30.18 & RG & 0.22 &  0.46 &  1.16 & 0.69 & no\\
4C+34.28A & RG & 0.41 &  0.55 &  1.77 & 0.87 & no\\
4C+34.42 * & RG & 0.40 &  0.77 &  2.57 & 0.89 & yes\\
4C+39.32 * & RG & 0.36 &  0.79 &  1.98 & 0.68 & no\\
4C+39.35 & RG & 0.25 &  0.63 &  1.61 & 0.69 & no\\
4C+40.11 & RG & 0.20 &  1.64 &  4.46 & 0.75 & no\\
4C+46.32 * & RG & 0.40 &  0.48 &  2.01 & 1.07 & no\\
4C+47.37 * & RG & 0.42 &  0.79 &  1.95 & 0.67 & no\\
4C+48.25 & RG & 0.45 &  0.79 &  2.32 & 0.80 & no\\
4C-01.59 & RG & 0.27 &  0.56 &  1.29 & 0.62 & no\\
4C38.35 & RG & 0.47 &  0.54 &  2.01 & 0.98 & yes\\
B21325+32 & RG & 0.26 &  1.51 &  3.33 & 0.59 & no\\
B22347+30 & RG & 0.37 &  0.40 &  1.10 & 0.76 & no\\
MRC1519+108 & RG & 0.20 &  0.51 &  1.44 & 0.77 & no\\
PKS0054+018 * & RG & 0.29 &  0.43 &  1.26 & 0.81 & no\\
SDSSJ090320.45+523336.1 & RG & 0.31 &  0.31 &  1.14 & 0.96 & no\\
SDSSJ113313.17+500840.0 & RG & 0.31 &  0.83 &  2.36 & 0.78 & no\\
SDSSJ22163 & RG & 0.37 &  0.25 &  1.16 & 1.14 & no\\
\enddata
\tablerefs{Redshifts and radio luminosities are extracted from NED database}\\
{\footnotesize Objects with asterisks (*) beside the names are part of the re-matched sample}
\label{wholesample}
\end{deluxetable}

\begin{deluxetable}{lccccccc}
\tablecaption{LRIS Sample}
\tabletypesize{\scriptsize}
\tablecolumns{8}
\tablehead{\colhead{Object Name} & \colhead{Object Type}  & \colhead{Redshift}  & \colhead{F(1400 MHz)}  & \colhead{F(365 MHz)} & \colhead{Spectral Index} & \colhead{EELR Detected?} & \colhead{Slit Width} \\
\colhead{} & \colhead{} & \colhead{} & \colhead{Jy} & \colhead{Jy} & \colhead{} & \colhead{} & \colhead{arcsec}}
\startdata
3C381 & RG & 0.16 &  3.88 &  8.71 & 0.60 & yes & 8.7\\
TXS1821+640 * & RG & 0.20 &  0.34 &  0.99 & 0.79 & no & 8.7\\
TXS1732-092 * & RG & 0.37 &  2.20 &  5.56 & 0.69 & yes & 8.7\\
4C-03.72 & RG & 0.19 &  0.52 &  1.69 & 0.87 & yes & 8.7\\
TXS2158+048 * & RG & 0.23 &  0.26 &  0.83 & 0.85 & yes & 8.7\\
3C436 & RG & 0.21 &  3.40 & 10.25 & 0.82 & yes & 8.7\\
B2 2225+24 & RG & 0.32 &  0.23 &  -- & -- & yes & 8.7\\
4C+39.72 * & RG & 0.21 &  0.81 &  2.10 & 0.71 & yes & 8.7\\
4C+03.56 & RG & 0.15 &  0.54 &  2.12 & 1.02 & no & 8.7\\
3C458 * & RG & 0.29 &  2.75 &  5.66 & 0.54 & yes & 8.7\\
3C459 & RG & 0.22 &  4.25 & 17.21 & 1.04 & yes & 8.7\\
3C460 & RG & 0.27 &  1.57 &  5.44 & 0.93 & yes & 8.7\\
3C462 * & RG & 0.39 &  2.46 &  6.46 & 0.72 & yes & 8.7\\
4C+45.02 & RG & 0.37 &  0.45 &  1.91 & 1.07 & no & 8.7\\
TXS0032+301 & RG & 0.17 &  0.27 &  1.09 & 1.04 & no & 8.7\\
3C020 & RG & 0.17 & 11.53 & 30.30 & 0.72 & no & 8.7\\
3C042 * & RG & 0.40 &  2.88 &  8.92 & 0.84 & no & 8.7\\
3C018 & RG & 0.19 &  4.60 & 11.33 & 0.67 & yes & 8.7\\
3C026 & RG & 0.21 &  2.21 &  7.62 & 0.92 & yes & 8.7\\
TXS0111-002 * & RG & 0.39 &  0.34 &  0.93 & 0.75 & no & 8.7\\
3C063 & RG & 0.17 &  3.42 & 12.52 & 0.97 & no & 8.7\\
3C064 * & RG & 0.27 &  2.53 &  4.63 & 0.45 & yes & 8.7\\
3C067 & RG & 0.31 &  3.07 &  8.01 & 0.71 & yes & 8.7\\
3C073 & RG & 0.20 &  2.06 &  5.80 & 0.77 & no & 8.7\\
3C079 & RG & 0.26 &  4.47 & 13.30 & 0.81 & yes & 8.7\\
TXS1118+315 & RG & 0.33 &  0.18 &  0.59 & 0.88 & no & 1.5\\
4C29.44 * & RG & 0.33 &  1.64 &  5.34 & 0.88 & yes & 1.5\\
3C268.2 * & RG & 0.36 &  1.47 &  2.89 & 0.50 & yes & 1.5\\
3C284 & RG & 0.24 &  1.92 &  7.07 & 0.97 & yes & 1.5\\
TXS1311+321 & RG & 0.30 &  0.14 &  0.33 & 0.67 & yes & 1.5\\
TXS1331+381 & RG & 0.38 &  0.16 &  0.55 & 0.93 & no & 1.5\\
4C39.41 & RG & 0.25 &  0.34 &  1.54 & 1.12 & no & 1.5\\
3C299 * & RG & 0.37 &  3.06 &  8.05 & 0.72 & yes & 1.5\\
TXS1426+394 & RG & 0.26 &  0.25 &  0.45 & 0.43 & no & 1.5\\
3C300 & RG & 0.27 &  3.60 & 10.47 & 0.79 & yes & 1.5\\
\enddata
\tablerefs{Redshifts and radio luminosities are extracted from NED database}
{\footnotesize Objects with asterisks (*) beside the names are part of the re-matched sample}
\label{lrisobs}
\end{deluxetable}

\section{Data Analysis}

\subsection{SNIFS Data}

Data reduction was mostly handled by the existing SNIFS pipeline \citep{2001MNRAS.326...23B,2006ApJ...650..510A}, which includes the standard processes of bias subtraction, flat-fielding, wavelength and flux calibration, and sky subtraction. EELR detections were determined by scaling and subtracting a continuum image from a corresponding emission-line image. The emission-line images are sums of all slices with significant [\ion{O}{3}]~$\lambda 5007$ emission. For objects without nuclear [\ion{O}{3}] detections, emission-line images were made by summing up 35 slices of the data cube centered on the wavelength where the [\ion{O}{3}] line should be according to the object's redshift. The continuum images were made by taking the median of a hundred continuum slices on each side of the [\ion{O}{3}] and H$\beta$ emission line region. The scaling factor was computed to minimize the residual in the inner 2 pixels radius centered on the continuum peak. If a significant [\ion{O}{3}] residual was found in the continuum subtracted images through visual inspection, we considered those to be detections. Detected EELRs and their physical properties are shown in Table \ref{eelrsample}. 

The nuclear emission for each object was calculated by summing up the [\ion{O}{3}] emission in the 2\arcsec~ diameter aperture around the continuum peak. The 2\arcsec~ aperture is slightly bigger than the average PSF, which varies from night to night but has an average of $1\farcs5$. This aperture is selected to account for some of the scattered nuclear emission without including a significant amount of extended emission. The EELR luminosity was calculated from the continuum subtracted images. 
There are 5 EELR quasars, PKS 1048-090, 4C+11.50, TXS1745+163, 4C\,29.02, and 3C\,323.1, that overlap with the \citet{1987ApJ...316..584S} sample. All of the nuclear [\ion{O}{3}] measurements agree within 0.15 dex despite different methods of observation and calibration. Three objects, however, have significantly higher extended [\ion{O}{3}] ($\sim 0.5$ dex) in the SNIFS measurements. This may be because \citet{1987ApJ...316..584S} excluded the inner 4\arcsec~ diameter from the EELR measurements to avoid quasar scattered light, which can be scaled and subtracted in the SNIFS data cube.  

The radio maps for the objects with EELR detections are compiled from the sources cited in Table \ref{eelrsample}. Most sources, as expected, have double lobe morphologies and the radio axis PAs can be easily calculated. Some objects do not have well resolved radio maps and their radio PAs are left blank. For comparison, the PA of the EELR elongation axes are estimated. While many EELRs appear to have a preferred direction of extension, some appear round and do not have well defined elongation axes. 

\subsection{LRIS Data}

The spectra were background subtracted, wavelength-calibrated, flux-calibrated and extracted using standard IRAF routines. For each object we extracted a nuclear spectrum using a 2\arcsec~ wide aperture centered on the continuum peak and an extended spectrum with a wider aperture scaled to include any remaining emission from the object. The nuclear [\ion{O}{3}] $\lambda 5007$ emission was calculated by fitting a Gaussian profile to the [\ion{O}{3}] line of the nuclear spectrum. The luminosity of the extended emission was calculated by first subtracting the nuclear spectrum from the extended spectrum, then fitting a Gaussian profile to the nuclear-subtracted [\ion{O}{3}] line. 

There are a 4 overlapping objects between the LRIS and the SNIFS observations (3C\,299, 3C\,284, 3C\,456, and 3C\,459), and it is important to note that there appear to be some discrepancies between measurements of the two instruments. The LRIS values for the objects observed with the $1\farcs5$ slits (3C\,299, 3C\,284) gave nuclear and extended [\ion{O}{3}] luminosities slightly lower than the SNIFS observations, but the values agree within 0.3 dex (factor of 2). The missing flux in the LRIS measurements may have landed outside the slit. 

The objects observed with the $8\farcs7$ slits (3C\,456, 3C\,459) appear to give nuclear [\ion{O}{3}] luminosities significantly higher than the SNIFS values ($\sim 0.7$ dex). The extended [\ion{O}{3}] measurements are also higher than SNIFS in this case, but they are in better agreement ($< 0.4$ dex). This may be due the the inclusion of extended emission in the `nuclear' apertures with a wider slit, and also inclusion of more emission in general due to a slit wider than the SNIFS field of view. Because of dome seeing and varying weather conditions, the SNIFS sample on average has seeing that is larger and also more variable than the LRIS data set, which may also contribute to some discrepancies. Unfortunately, because long slit observations do not give enough information on the EELR and host galaxy morphologies, the discrepancies due to instrument configurations are difficult to calculate for individual objects. Therefore the LRIS fluxes given in this paper should be treated as order of magnitude estimates instead of accurate values. 
 
Note: Because of the different nature of the observations, comparisons including the LRIS data will be restricted to the detection rates, and total nuclear and extended [\ion{O}{3}] luminosities. The long-slit data does not contain information about resolved 2-D kinematics and morphologies. Most LRIS observations use a very wide $8.7\arcsec$ slit, so the spectral resolution is too low to extract kinematic information comparable to the SNIFS data. Although some information about the EELR linear extents can be found in the LRIS data, the maximum extent of an EELR is highly dependent on the slit orientation, and any measurements from long-slit data will likely be underestimates. Therefore, we do not attempt to compare the EELR sizes from the LRIS data to those from the SNIFS data.

\subsection{Re-matching samples}

As discussed in \S\ \ref{sec:sample}, not all powerful steep-spectrum radio galaxies harbor a hidden luminous quasar. We eliminated objects that may be LERGs by keeping only the objects with a well detected [\ion{O}{3}] line with equivalent width $> 10$\AA. Because our SNIFS observations are quite shallow, $\sim 30\%$ of the observed RGs may have an [\ion{O}{3}] line with EW$ > 10$\AA~ that nevertheless is not detected. Another $\sim 20\%$ are confirmed to be LERGs. In the LRIS RG sample, $\sim 10\%$ are shown to be LERGs. {\it All discussion from this point on refers only to the remaining RG sample after the possible LERGs are eliminated.} As expected, no LERGs show any extended emission.

\begin{figure*}[t]
\centering
\includegraphics[width=6.in]{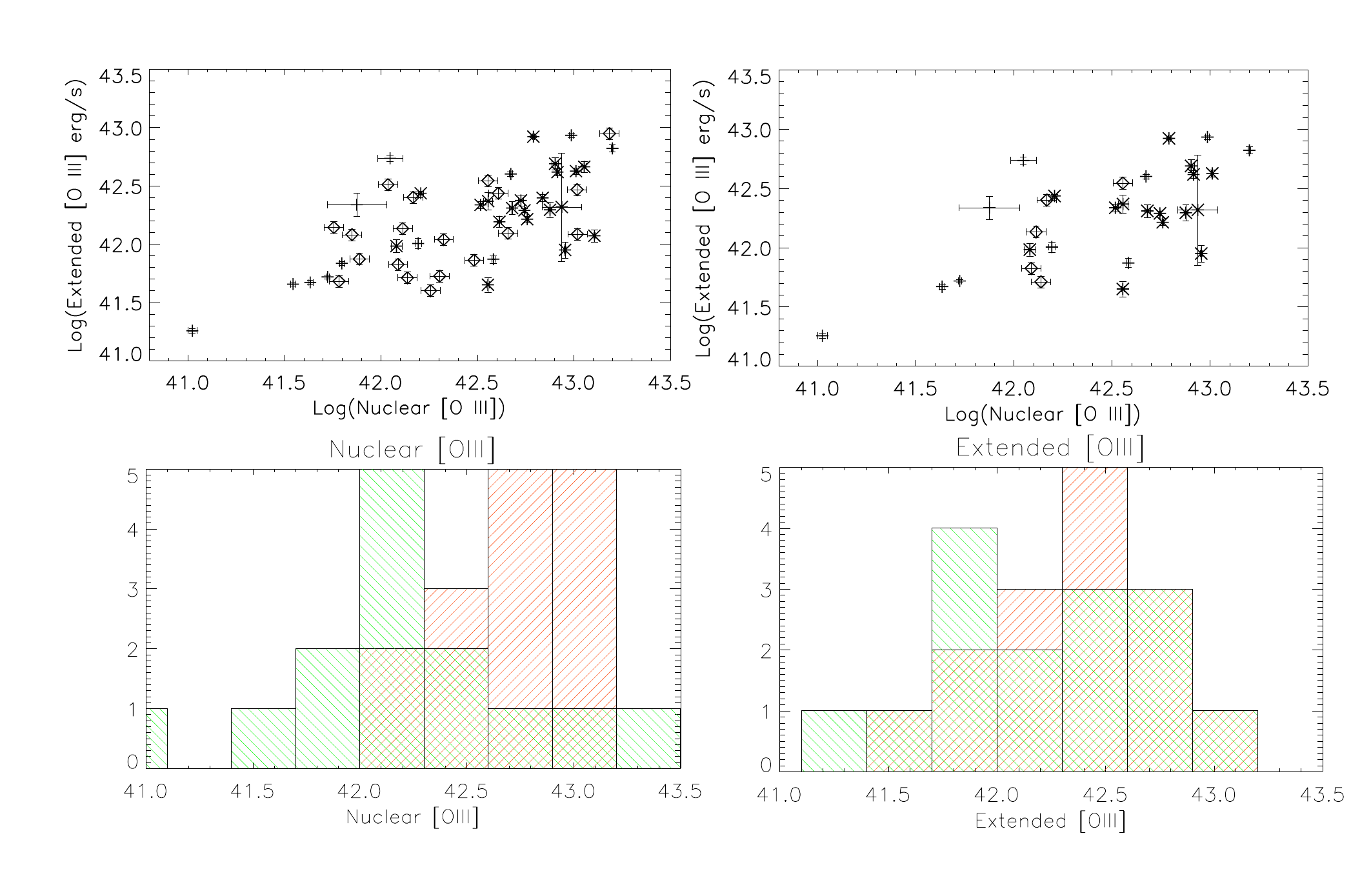}
\caption{Nuclear [\ion{O}{3}] vs. Extended [\ion{O}{3}] luminosity. Stars represent QSOs, crosses represent RGs and diamonds represent LRIS RG observations. Top Left: All detected EELRs. Top Right: Only the matched sample. Bottom Left: Nuclear [\ion{O}{3}] distribution of the matched sample. The red bars represent the quasars and the green bars represent the RGs. Bottom Right: Extended [\ion{O}{3}] distribution of the matched sample with EELR detections.}
\label{exo3plot}
\end{figure*}

When the LERGs are excluded, the SNIFS sample is no longer as well matched in redshifts and radio properties. Because the [\ion{O}{3}] luminosities are known to correlate with radio power, and both are subject to selection effects with respect to redshift, we need to re-select the sample to check for any biases that may have been caused by the deficit of RGs. With the combined (LRIS and SNIFS) sample of HERGs, we rematched each one of them to a quasar that has a similar redshift, radio power and spectral index. Members of the rematched sample are indicated by asterisks beside the object names in Table \ref{wholesample} and \ref{lrisobs}. A comparison between the [\ion{O}{3}] properties of the entire sample and the re-matched sample is shown in Figure \ref{exo3plot}. From the top 2 panels in Figure \ref{exo3plot}, we see that the [\ion{O}{3}] distribution of the re-matched sample appears to be representative of the entire sample. Therefore we include the entire SNIFS sample in the subsequent discussion about morphologies and kinematic properties.

\subsection{Radio Selection Effect and Detection Rates}

\begin{figure*}[t]
\centering
\includegraphics[width=6.in]{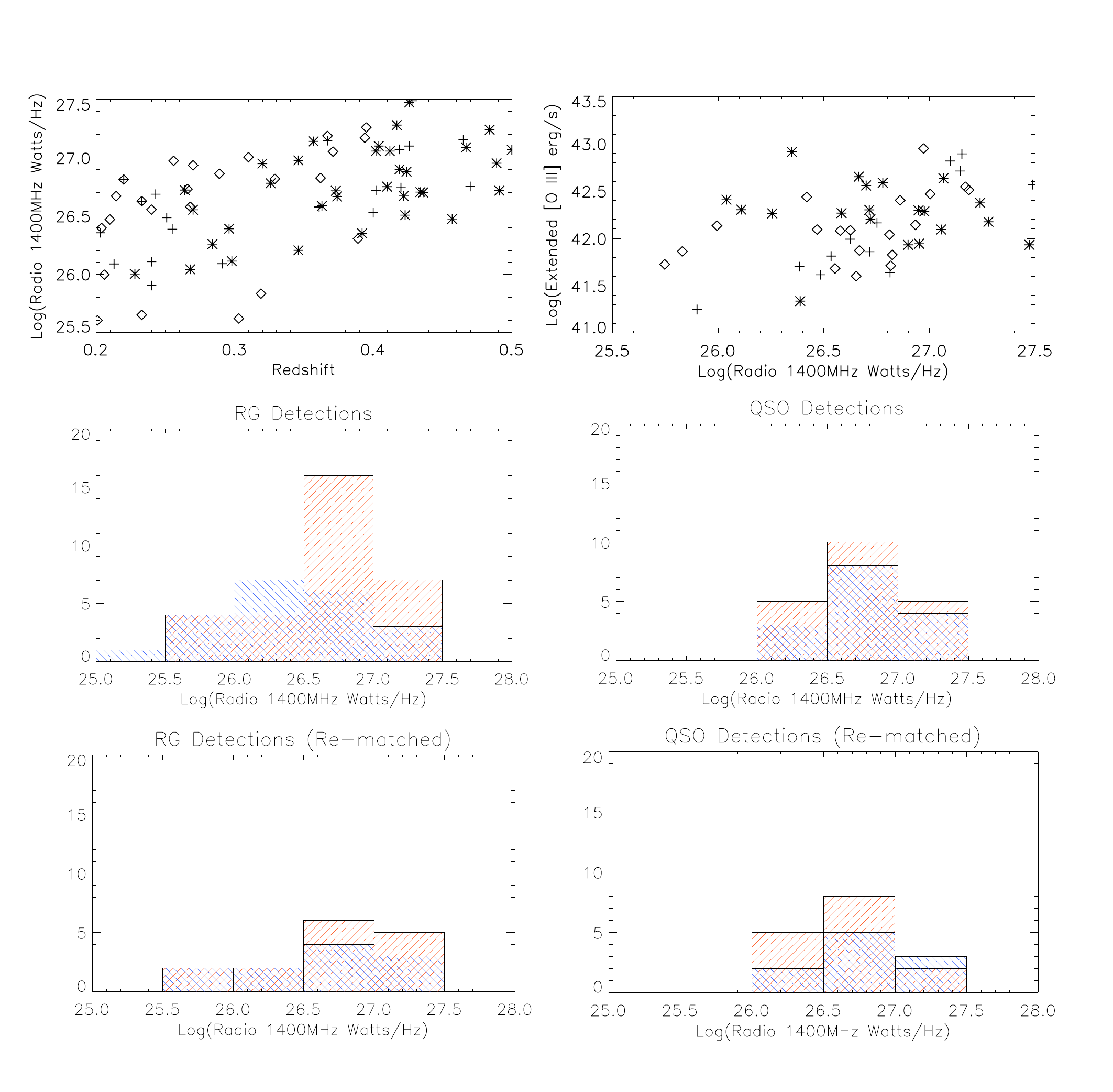}
\caption{Top left: Radio selection effect; higher radio power objects are selected at higher redshifts. Top right: Extended [\ion{O}{3}] luminosities as a function of radio power. Asterisks - quasars, crosses - RGs, diamonds - LRIS RG observations. Middle row: Detection histogram in different radio power bins for all RGs (HERGs) and QSOs. Bottom row: Same as middle row but for the re-matched sample. Red - objects with EELR detection, Blue - objects with no EELR detection.}
\label{radiodet}
\end{figure*}

The samples are selected to be above 1 Jy based on the TXS 365 MHz catalog. Because there is a range of redshifts in our sample, the radio survey is actually probing different rest frame frequencies for each object. We estimate the rest frame 365 MHz and 1400 MHz fluxes for each object by using the redshift and spectral index. We then convert the fluxes to absolute radio luminosities (Watts Hz$^{-1}$). As shown in Figure \ref{radiodet}, because our selection is based on observed radio luminosity, the higher redshift sources have higher radio powers. There appears to be a slight positive correlation between the extended [\ion{O}{3}] luminosity and the radio power, but that is due to the selection of more luminous objects, at both optical and radio wavelengths, at higher redshifts. Also the physical field of view at higher redshift is larger, which means we can pick up more emission from large sources. 

The EELR detection rates, on the other hand, do not depend strongly on radio power within the range of radio power considered in this study. The middle 2 panels of Figure \ref{radiodet} show the detection rates among all the RGs (high-excitation) and the quasars observed with SNIFS and LRIS. The bottom 2 panels shows the detection rates in the re-matched sample. The detection rates in the re-matched sample appear to be representative of the entire sample. The total detection rates for each of the lower 4 panels are all $\sim 50 - 60\%$.

\section{Results}

\subsection{Detection Rates and [O III] Luminosities}

In the SNIFS observations, EELRs are clearly detected around 20 out of 39 quasars, 13 out of 20 HERGs, and 0 out of 21 LERGs. In the LRIS RG sample, EELRs are detected around 19 out of 31 objects. The LRIS observations are deeper than the SNIFS data. To remain consistent, we do not consider detections from LRIS that are too faint for SNIFS to detect. An EELR is considered detected in LRIS data if Log(extended [\ion{O}{3}] erg s$^{-1}$) $>  41.5$, above which the EELRs can be easily detected in the SNIFS observations. 

In the rematched sample, the detection rate of the EELRs is the same in the RG and the quasar sample (15 out of 25). The nuclear [\ion{O}{3}] luminosities of quasars are higher than those of the RGs, possibly due to orientation effects such as partial obscuration of the narrow line regions in some RGs. .

\begin{landscape}
\begin{centering}
\begin{deluxetable}{lccccccccc}
\tablecaption{EELR Sample}
\tabletypesize{\scriptsize}
\tablecolumns{7}
\tablehead{\colhead{Object} & \colhead{Object} & \colhead{Nuclear}& \colhead{Extended} & \colhead{Radio} & \colhead{EELR} & \colhead{EELR} & \colhead{EELR} & \colhead{EELR}  & \colhead{EELR}\\
\colhead{Name} & \colhead{Type} & \colhead{[OIII]}& \colhead{[OIII]} & \colhead{PA} & \colhead{PA} & \colhead{Area} & \colhead{Extent}  & \colhead{FWHM}  & \colhead{Mass}\\
\colhead{} & \colhead{} & \colhead{Log(erg/s)} & \colhead{Log(erg/s)} & \colhead{} & \colhead{} & \colhead{kpc$^2$} & \colhead{kpc}  & \colhead{km/s} & \colhead{Log(M$_\odot$)}}
\startdata
3C047 & QSO & 42.96 & 41.95 &  35(1) & -- &  123.2 &   19.1  &  751.0 &  8.1\\
4C+21.26 & QSO & 42.56 & 41.65 &  79(2) & -- &   95.0 &   16.1 &  441.6 &  7.8\\
4C+41.04 & QSO & 43.05 & 42.66 &  72(3) & -- &  374.5 &   26.3 &  274.0 &  8.9\\
4C+25.40 & QSO & 42.21 & 42.44 &  29(4) & 59 &  149.7 &   22.8 &  388.5 &  8.6\\
4C+25.01 & QSO & 42.74 & 42.29 & -26(5) & -- &  119.0 &   13.0 &  553.6 &  8.5\\
4C+29.02 & QSO & 42.56 & 42.37 &  45(1) & -- &  234.4 &   24.9 &  488.1 &  8.6\\
4C+15.09 & QSO & 43.11 & 42.07 &   0  & -- &  138.7 &   23.8 & 500.0 &  8.3\\
4C+02.23 & QSO & 42.68 & 42.31 & -27(6) & -- &  138.6 &   18.4 &  352.2 &  8.5\\
4C+09.35 & QSO & 42.52 & 42.34 & 143(4) & 135 &  127.5 &   16.3 &  112.0 &   8.5\\
4C+22.25 & QSO & 42.08 & 41.98 &  30(4) & -- &  145.1 &   19.9 &  406.0 &  8.2\\
4C+31.38 & QSO & 42.61 & 42.19 &  45(8) &$\sim -45$ &  130.1 &   18.1 &  643.2 &  8.4\\
TXS1220+373 & QSO & 42.73 & 42.37 &  90(9) & -- &  278.8 &   25.6 &  466.8 &  8.6\\
4C+11.50 & QSO & 43.01 & 42.63 & -22  & -- &  177.5 &   18.2 &  237.9 &   8.8\\
4C+27.38 & QSO & 42.90 & 42.69 &   0(6) & 90 &  163.7 &   21.4 &  445.4 &  8.9\\
TXS1745+163 & QSO & 42.79 & 42.92 &   0  & -60 &  174.2 &   19.5 &  604.9 &  9.1\\
4C+08.64 & QSO & 42.84 & 42.40 &   0(10) & 30 &  154.1 &   19.3 &  283.9 &  8.6\\
4C+11.72 & QSO & 42.92 & 42.62 & -51(6) & -49 &  196.4 &   18.8 &  520.3 &   8.8\\
3C246 & QSO & 42.88 & 42.29 & -63  & $\sim -20$ &  148.2 &   18.5 &  392.3 &   8.5\\
3C323.1 & QSO & 42.76 & 42.21 &  20  & -65 &  130.6 &   18.1 &  165.5 &  8.4\\
3C277.1 & QSO & 42.94 & 42.32 & -45(12) & -45 &  133.4 &   15.3 &  449.3 &  8.5\\
3C099 & RG & 43.20 & 42.82 &  45(13) & 45 &  187.3 &   19.1 &  808.5 &  9.0\\
3C244.1 & RG & 42.67 & 42.60 & -15(13) & -45 &  199.1 &   22.3 &  512.7 &  8.8\\
3C284 & RG & 41.80 & 41.83 & -80(14) & 67 &   68.0 &   14.8 &  241.4 &  8.0\\
3C299 & RG & 42.05 & 42.74 &  68(15) & -- &  206.3 &   16.8 &  758.2 &  8.9\\
3C456 & RG & 42.19 & 42.01 &  18(16) & 90 &   64.9 &   15.03 &  793.5 &  8.2\\
3C459 & RG & 41.54 & 41.66 &  90(16) & -60 &   49.7 &    15.5 &  517.4 &  7.8\\
4C+09.44 & RG & 41.63 & 41.67 &  62(2) & 45 &  101.9 &   17.8 &    100.0 &  7.9\\
4C+10.71 & RG & 41.72 & 41.72 &  56(2) & 90 &   48.2 &   11.0 & 569.3* &   7.9\\
4C+20.27 & RG & 42.99 & 42.93 &   0(2) & 45 &  236.9 &   24.3 &  267.9 &  9.1\\
4C+22.45 & RG & 42.21 & 42.98 &   0(2) & -63 &  191.7 &   22.6 &    200.0* &  9.25\\
4C+34.42 & RG & 42.58 & 41.87 &  23(17) & 45 &   69.1 &   13.1 &  182.2 &  8.1\\
4C+38.35 & RG & 41.87 & 42.34 &  24(9) & 26 &  45.0 &   16.7 &    200.0** &  8.5\\
PKS0230-027 & RG & 41.02 & 41.26 & -35(18) & -10 &   61.1 &   12.19 &    200.0* &  7.4\\
3C381 & RG & 42.61 & 42.44 &  --   &  --  &  --  &  --  &  --  &  8.7\\
4C-03.72 & RG & 42.31 & 41.72 &  --   &  --  &  --  &  --  &  --  &  8.0\\
3C436 & RG & 41.89 & 41.87 &  --   &  --  &  --  &  --  &  --  &  8.1\\
B2 2225+24 & RG & 42.48 & 41.86 &  --   &  --  &  --  &  --  &  --  &  8.1\\
4C+39.72 & RG & 42.11 & 42.13 &  --   &  --  &  --  &  --  &  --  &  8.4\\
3C456 & RG & 43.02 & 42.09 &  --   &  --  &  --  &  --  &  --  &  8.3\\
3C458 & RG & 42.17 & 42.40 &  --   &  --  &  --  &  --  &  --  &  8.6\\
3C459 & RG & 42.33 & 42.04 &  --   &  --  &  --  &  --  &  --  &  8.3\\
3C460 & RG & 41.85 & 42.08 &  --   &  --  &  --  &  --  &  --  &  8.3\\
3C462 & RG & 42.55 & 42.54 &  --   &  --  &  --  &  --  &  --  &  8.8\\
3C018 & RG & 42.26 & 41.60 &  --   &  --  &  --  &  --  &  --  &  7.8\\
3C026 & RG & 42.66 & 42.09 &  --   &  --  &  --  &  --  &  --  &  8.3\\
3C067 & RG & 43.02 & 42.47 &  --   &  --  &  --  &  --  &  --  &  8.7\\
3C079 & RG & 43.18 & 42.95 &  --   &  --  &  --  &  --  &  --  &  9.2\\
4C29.44 & RG & 42.14 & 41.71 &  --   &  --  &  --  &  --  &  --  &  7.9\\
3C268.2 & RG & 42.09 & 41.83 &  --   &  --  &  --  &  --  &  --  &  8.1\\
3C284 & RG & 41.78 & 41.68 &  --   &  --  &  --  &  --  &  --  &  7.9\\
3C299 & RG & 42.04 & 42.51 &  --   &  --  &  --  &  --  &  --  &  8.7\\
3C300 & RG & 41.76 & 42.14 &  --   &  --  &  --  &  --  &  --  &  8.4\\

\enddata
\tablerefs{(1) \citet{1984ApJ...283..515B} (2) FIRST Survey (3) \citet{1989AJ.....98..419V} (4) \citet{1984AJ.....89.1658G} (5) \citet{1985ApJ...299..799S} (6)\citet{1993ApJS...86..365P} (7) \citet{1998PASP..110..111H} (8) \citet{1989MNRAS.240..657S} (9) \citet{1982MNRAS.199..611A} (10) \citet{1983AJ.....88..709H} (11) \citet{1994AJ....108.1163K} (12) \citet{1991MNRAS.250..215A} (13) \citet{1995ApJS...99..349N} (14) \citet{1984MNRAS.210..929L} (15) \citet{1995A&AS..112..235A} (16) \citet{1998ApJS..119...25H} (17) \citet{1993A&AS..101..431B} (18) \citet{1986MNRAS.218...31D}}
\label{eelrsample}
\end{deluxetable}
\end{centering}
\end{landscape}

\subsection{EELR morphologies}

The morphology, velocity and velocity dispersion fields of detected EELRs are shown in Figure \ref{eelrmaps1}. The EELR morphologies encompass a wide range of shapes. Some appear to have a clear axis of extension, and the PA for those objects is measured from the EELR contour plots and noted in Table \ref{eelrsample}. Some extend out from the nucleus in almost all directions. These objects do not have measured PAs in Table \ref{eelrsample}. Some others, such as 3C\,244.1 and PKS\,0230-027 have separate clouds several kpc from the nucleus. The EELR PAs for these objects are determined by the relative positions between the detected EELR clouds and the nucleus.

\subsection{Radio Alignment}

Previous work with narrow-band imaging has shown that the radio structure does not necessarily correlate with the EELR structure in both quasars and radio galaxies \citep[e.g.,][]{2007ApJ...666..794F,2008ApJS..175..423P}. The lack of correlation shows that the EELR gas is not simply an outflow entrained in the radio jets, but is likely to have been ejected by the wide-solid-angle blast wave associated with the initiation of the radio jet. There also appears to be an evolutionary trend in radio-optical alignment with age. Younger ($10^{3}$ to $10^{5}$ years old) versions of FR II radio sources have extended-emission regions that are more likely to be closely aligned with the radio axes \citep{2000AJ....120.2284A,2008ApJS..175..423P,2013ApJ...772..138S}.

\begin{figure*}[t]
\centering
\includegraphics[angle=0, width=6.in]{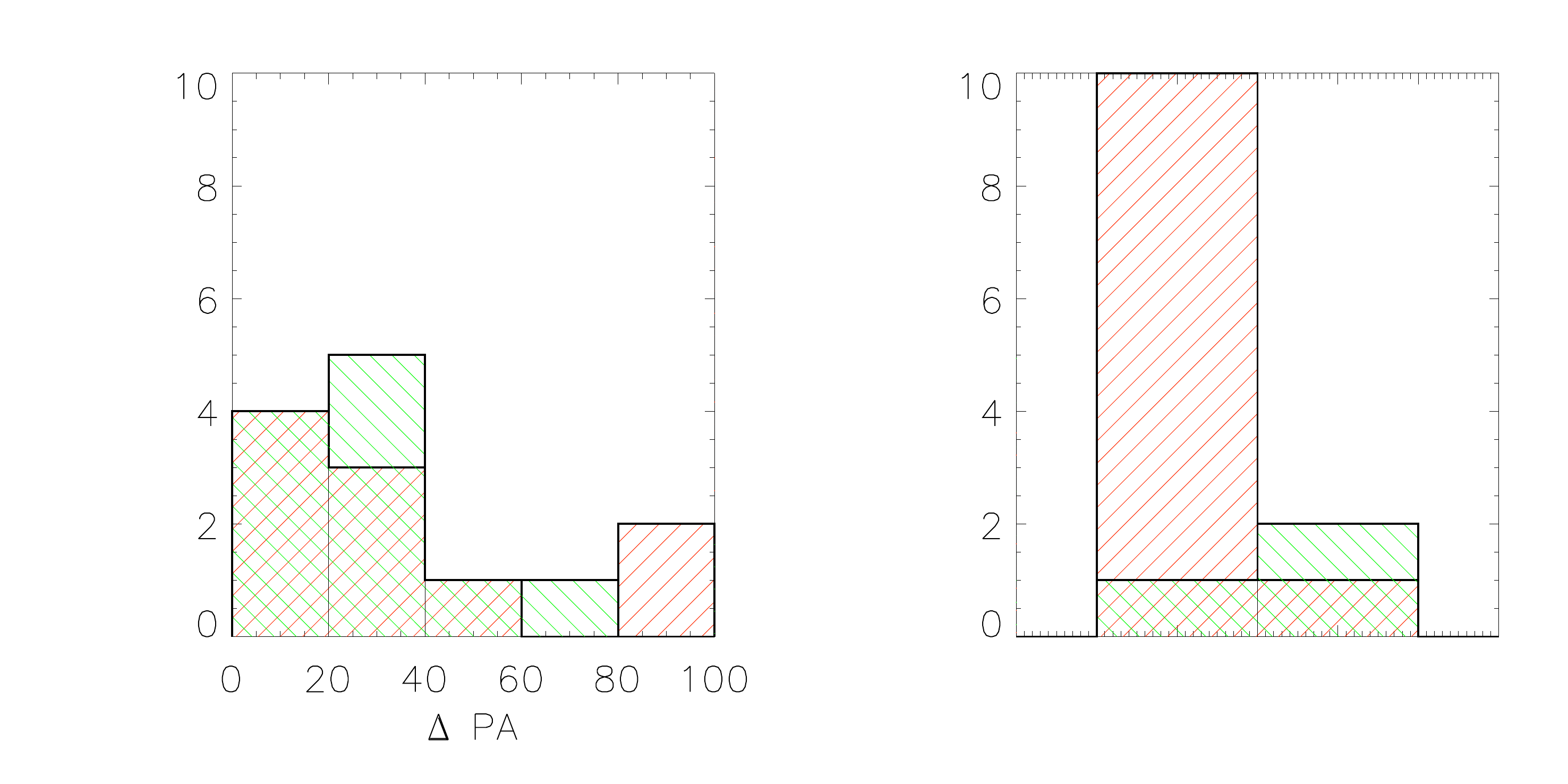}
\caption{Left: $\Delta$PAs between EELR PA and radio axis PA. Red represent quasars and green represent RGs. Right: left bar - EELRs that have no particular extension axis, right bar - objects with no well resolved radio maps. The same color scheme as plot on the left.}
\label{padiff}
\end{figure*}

\subsection{Velocity Dispersion}

The median velocity dispersion (FWHM) of most EELRs is between 350 and 600 km s$^{-1}$. There are a few cases of higher velocity dispersions with FWHM $> 1000$ km s$^{-1}$. The velocity dispersion field in most objects appears to be rather uniform with no large scale organized variations. The few exceptions, such as 3C\,99 and 4C\,31.38, appear to have clear velocity dispersion gradients and are also the objects that reach the highest velocity dispersions. The velocity dispersions are corrected for instrumental broadening, which is subtracted in quadrature from the measured FWHMs.

\subsection{Velocity Field}

There are a range of kinematic patterns among the luminous EELRs. Some have line of sight velocities fairly close to the systemic and do not show a detectable velocity gradient (e.g., 3C\,456). Some appear to have one well defined velocity gradient (e.g., TXS1741+297). Others show more complicated, but nevertheless organized velocity patterns. The velocity gradients do not appear to preferentially align with the radio axes or the direction of EELR elongations. 

In attempt to quantify the characteristics of the velocity field, we fit the velocity fields with simple models of two likely scenarios for EELRs: a rotating disk and a biconical outflow \citep[][and references therein]{2011ApJ...739...69M}.The rotating disk is modeled by concentric, coplanar rings of rotating gas with a flat rotation curve ($V(R) = V_o$). We have also considered the solid-body rotating disk $V(R) = kR$ and the Keplerian rotating disk $V(R) = kR^{-0.5}$. The solid-body disk produces equally good or slightly better fits in some cases, and significantly worse fits in others. The Keplerian disk results in worse fits in all cases. The flat rotation disk gives the best overall fits, so, for simplicity, we use it for all objects. The projected velocity along the line of sight follows the cosine law approximation $V(R,\psi) =  V_c(R)sin(i)cos(\psi)$, where i is inclination (i = 0 for a disk viewed face on), R is the radius and $\psi$ is the azimuthal angle measured from the projected major axis \citep{2006MNRAS.366..787K}. 

The calculated velocities are based on redshifts drawn from NED, which are not always precise. As a result, our velocities can be offset from the true velocities by up to a few hundred km s$^{-1}$. Our data are not deep enough to detect stellar absorption lines, which are needed to determine more precise systemic velocities. To compensate for the offsets from the true velocities, we include a velocity offset parameter in our velocity structure models. The free parameters for the rotating disk model are the inclination and PA of the disk, the circular velocity, and the velocity offset. 

The biconical outflow is modeled by two cones placed apex to apex. A pictorial illustration can be found in \citet{2011ApJ...739...69M}. The velocity of the gas in the cones scales linearly with the distance from the nucleus ($V(r) = V_o + kr$). The free parameters are: the opening angle, PA and inclination of the cones, the rate of the velocity change with respect to radius, and a velocity offset. The projected velocity at each point on the cone is $V_p = V(r)sin(\alpha)$ where $\alpha$ is the angle between the observer's line of sight and the direction of the velocity vector. The observed velocity is the average of all projected velocities along the line-of-sight. The gas velocity can increase, decrease, or stay constant with radius. The increasing velocity with radius assumes that the gas starts with relatively low velocity and accelerates as it received energy injected by the jets. Decreasing velocity represent gas that receives all of its energy close to the center and decelerates as it collides with material outside. Constant velocity gas, on the other hand, assumes no significant force on the gas after its ejection from the center. The cones can either be filled, or have cone walls 20\arcdeg~ thick. We have tested both cases on all the velocity fields and determined that, in all cases, the accelerating cone with 20\arcdeg~ wall thickness produces the closest fits.

In the SNIFS sample, there are 19 velocity fields with significant organized variation within the field. For each object, we perform three different kinds of disk/cone combination: (1) only the disk component, (2) only the bicone component, and (3) a linear combination of the disk and bicone components with the fractional contribution from each model being a free parameter. We used the MPFIT routine \citep{2009ASPC..411..251M} to optimize our fits. 
 
The fit results and fitted parameters are shown in Figure \ref{modelfit4} and Table \ref{fitparam}. The model with the least free parameters, the disk model, can produce fits without significant systematic residuals for eight velocity fields. In these cases, the biconical model results in worst fits than the disk model for all eight objects. There are three velocity fields that benefit from the combination of disk and cone components, and one that can be fitted by just the bicone model. Among the objects not listed in Table \ref{fitparam}, five objects cannot be well described by the simple models considered here (4C\,25.40, 4C\,29.02, 4C\,15.09, 3C\,459, 3C\,99), two objects cannot be fitted by the disk or bicone alone, and they do not have enough spaxels to constrain the combined model (4C\,34.42, 4C\,25.01). The remaining objects do not show significant velocity variation within our field of view. 

Theoretically, if there is a biconal outflow with high inclination (pointed along our line of sight), the overlapping cones will be seen as multiple Gaussian components in the line profile. Where there is a combination of disk and bicone components, we would also expect the line profiles to have multiple components. In practice, however, although some of our objects do have line profiles that slightly deviate from a single Gaussian, given the shallow and low-resolution nature of our data, the multi-Gaussian models are poorly constrained. Therefore, we do not attempt to distinguish emission that possibly come from different parts of the bicone, or different kinematic components. We also acknowledge the possibility that additional information on the undetected lower luminosity EELRs clouds and/or the EELRs outside of our field of view may change the results of the velocity model fit.

\begin{deluxetable}{lcccc}
\tablecaption{Fit Parameters}
\tablewidth{0pt}
\tabletypesize{\scriptsize}
\tablecolumns{5}
\tablehead{\colhead{Object Name} & \colhead{Disk Inclination} & \colhead{Disk PA} & \colhead{Disk Velocity (km/s)}& \colhead{Disk Fraction} \\ 
\colhead{} & \colhead{Cone Opening Angle} & \colhead{Cone Inclination} & \colhead{Cone PA} & \colhead{Velocity Offset}}
\startdata
3C 277.1 & $18.04\pm 4.90$ & $61.01\pm 3.77$ & $430.02\pm288.36$ & 1.00 \\ 
 & -- & -- & -- & $-1.8\pm0.6$ \\ 
4C 27.38 & $17.37\pm 3.26$ & $-47.58\pm 1.83$ & $296.35\pm231.28$ & 1.00 \\ 
 & -- & -- & -- & $-190.4\pm0.7$ \\ 
4C 21.26 & $10.71\pm 5.17$ & $15.70\pm 3.94$ & $376.60\pm127.06$ & 1.00 \\ 
 & -- & -- & -- & $-165.5\pm3.2$ \\ 
4C 02.23 & $18.27\pm 7.61$ & $-88.38\pm 2.70$ & $236.24\pm91.62$ & 1.00 \\ 
 & -- & -- & -- & $-105.1\pm2.7$ \\ 
4C 09.35 & -- & -- & -- & 0.00 \\ 
 & $25.00\pm 6.99$ & $83.58\pm 1.85$ & $-89.93\pm 2.76$ & $-117.7\pm5.6$ \\ 
4C 31.38 & $35.83\pm 3.96$ & $-31.26\pm 0.92$ & $162.35\pm16.18$ & 1.00 \\ 
 & -- & -- & -- &  $-61.6\pm0.5$ \\ 
TXS 1745+163 & $ 6.43\pm 4.34$ & $-162.95\pm 0.45$ & $2427.20\pm499.57$ & 1.00 \\ 
 & -- & -- & -- & $-74.9\pm1.0$ \\ 
4C 08.64 & $48.71\pm11.93$ & $-180.00\pm 2.80$ & $747.18\pm201.69$ & 0.14 \\ 
 & $49.01\pm 8.04$ & $ 3.69\pm 2.79$ & $52.86\pm 8.02$ & $-80.8\pm68.8$\\ 
4C 11.72 & $19.50\pm 4.43$ & $80.70\pm 3.69$ & $777.45\pm382.94$ & 0.78 \\ 
 & $41.85\pm 8.23$ & $87.92\pm 3.04$ & $-90.14\pm 7.26$ & $-172.4\pm35.6$ \\ 
4C 10.71 & $67.34\pm 2.06$ & $106.28\pm 1.38$ & $245.01\pm14.27$ & 1.00 \\ 
 & -- & -- & -- & $-161.6\pm2.7$ \\ 
3C 323.1 & $26.20\pm 7.25$ & $68.20\pm 6.43$ & $481.48\pm55.58$ & 0.74 \\ 
 & $61.29\pm 5.68$ & $ 3.83\pm 2.57$ & $-169.66\pm 2.69$ & $-51.73\pm43.67$ \\ 
3C299 & $22.84\pm 4.51$ & $158.26\pm 1.23$ & $563.08\pm91.76$ & 1.00 \\ 
 & -- & -- & -- & $-60.5\pm1.7$ \\ 
\enddata
\label{fitparam}
\end{deluxetable}

\subsection{EELR Mass}

We can estimate the EELR mass if we know the H$\beta$ flux and the gas density. We cannot get direct measurements of H$\beta$ fluxes from our data, but we can get a reasonable estimate using the [\ion{O}{3}] $\lambda 5007$ flux. Typical EELR clouds have [\ion{O}{3}]$\lambda 5007$ / H$\beta$ ratios ranging from $8 - 12$ \citep[e.g.,][]{2006ApJ...650...80F,2009ApJ...690..953F}. For our order of magnitude estimate we'll assume a [\ion{O}{3}] / H$\beta$ ratio of 10. The mass of the \ion{H}{2} region is expressed by:

\begin{equation}
\begin{centering}
M_H = \frac{4 \pi m_{p} f_{H\beta} d_{L}^2}{\alpha_{H\beta} n_{e} h \nu}
\end{centering}
\end{equation}

where $m_{p}$ is the proton mass, $d_{L}$ is the luminosity distance, $\alpha_{H\beta}$ is the effective recombination coefficient of H$\beta$ and $h\nu$ is the energy of a H$\beta$ photon. For $n_{e}$, the electron density, we assume a two-phase medium as has been observed in the EELR around 4C\,37.43 \citep[][see discussion in section 3.4]{2002ApJ...572..735S}: a density bounded component with $n_e \sim 2$ cm$^{-3}$ and an ionization bounded component with $n_e \sim 500$ cm$^{-3}$. We assume that each density component contributes half of the H$\beta$ flux. The estimated masses are listed in Table \ref{eelrsample}. The EELR masses range from $3.1 \times 10^7$ to $1.5 \times 10^9$ M$_{\odot}$. The median mass is $3.6 \times 10^8$ M$_{\odot}$, with a standard deviation of $\sim 0.4$ dex. Given a typical FR II lifetime of $\sim 10^7$ yrs \citep{2000AJ....119.1111B, 2008ApJ...676..147B}, the average mass outflow rate is $\sim 30$ M$_{\odot}$/yr. For the EELRs with masses reaching up to $10^9$ M$_{\odot}$, the outflow rate can be as high as $\sim 100$ M$_{\odot}$/yr.


\section{Discussion}
\label{sec:discussion}

\subsection{Morphology}
Previous studies have shown that the morphologies of the EELRs are globally chaotic. This is in agreement with the findings in our larger sample. The lack of morphological alignment between the optical structure and the radio axis does not necessarily mean that the radio source is not responsible for shaping the EELRs. Simulations have shown that the onset of radio jet is likely accompanied by the creation of an adiabatically expanding spherical bubble which may also be a driving source of the outflow \citep{2007Ap&SS.311..293S}. 

Compared with those of younger radio sources, the EELRs around these more evolved sources tend to be somewhat larger. The EELRs around young compact-steep-spectrum (CSS) sources studied in \citet{2013ApJ...772..138S} have linear extents of $< 15$ kpc, and an average size of $\sim 11.5$. The EELRs found in this study are mostly $> 15$ kpc and the average size is 18.5 kpc. It should be noted that our data has a limited field of view, so we may be underestimating the EELR sizes. While the optical size evolution is not nearly as dramatic as the radio size evolution, which spans almost two orders of magnitude, there does appear to be an increasing trend in size. 

The $\Delta$PAs for all detected EELRs are plotted in Figure \ref{padiff}. For the EELRs that have well defined PAs, the $\Delta$PAs are more likely to stay within $\lesssim 40$\arcdeg. However, any $\Delta$PA is possible. Those objects with no particular optical PAs are predominantly quasars. All except for one RG have EELRs with well defined axes of elongation. This difference between the RG and the quasar samples suggest that the 3-D morphology of the EELRs differs from a spherically symmetric distribution. A wide-angle bi-cone is more consistent with the observations. In this scenario, one would expect to see more circularly distributed EELRs when the cones are more aligned with our line of sight.

The EELRs around younger CSS sources have optical PAs that are typically aligned within $< 20$ degree of the radio axes. Comparing the sources at different evolutionary stage, there is a trend toward more misalignment between the optical elongation and the radio axes.

\subsection{Detection Rates and [\ion{O}{3}] Luminosities}
\label{discussion:detrate}
No significant differences are found in the detection rates and extended [\ion{O}{3}] luminosity of the quasar and radio galaxy sample. The consistent detection rates between the two subsample supports the unified model in which radio galaxies and quasar are the same objects with different relative orientations. This indicates that the EELRs are not preferentially obscured in any direction, unlike the nuclear [\ion{O}{3}] emission which is, on average, clearly weaker in the radio galaxy sample. Observations suggest that the narrow line regions of Seyfert 2 galaxies likely suffer more extinction than those of Seyfert 1s \citep[e.g.,][]{2011ApJ...727..130K}. The same orientation-dependent obscuration of the NLR is likely to be at work in these 
radio loud sources.

Above our radio luminosity cutoff, the detection rate and luminosities of EELRs do not appear to depend on radio loudness. More powerful radio jets are not more likely to produce EELRs, nor do they produce more luminous EELRs. However, all of our objects are fairly powerful radio sources, and it appears that luminous EELRs are seldom associated with radio-quiet objects \citep{1987ApJ...316..584S}, so this lack of correlation cannot continue to very low radio powers. 

The luminosities of EELRs appear to correlate, somewhat loosely, only with the nuclear [\ion{O}{3}] luminosities. The host galaxies of powerful radio sources are usually associated with giant ellipticals \citep[e.g.,][]{1996ApJ...473..713Z}, and the large mass of gas required to produce an EELR is likely acquired when a gas-rich late-type galaxy merged with the gas-poor elliptical. The brightness of the nuclear [\ion{O}{3}] is likely correlated with the overall amount of gas that merged into the elliptical. In this case brighter nuclear [\ion{O}{3}] may correlate with more gas being available for outflow/ionization throughout the galaxy.

\subsection{Velocity Dispersion}
The velocity dispersion maps of most EELRs do not show clear correlations with the radio structure. Data on the EELR around 3C\,249.1 in \citet{2006ApJ...650...80F} showed broad emission-line regions along the direction of the radio jet, possibly indicating disturbance from the radio jet. The only objects for which there is sign of possible disturbance from the jets in our sample are 3C\,99 and 4C\,31.38, where there are regions of high velocity dispersion ($FWHM \sim 1000$ km s$^{-1}$) along the radio jet paths. These two objects have the smallest corresponding radio structures, 6\arcsec and 2\farcs3 respectively, out of all objects with EELRs. 4C\,31.38 is considered a compact-steep-spectrum (CSS) source, which is a young \citep[$\sim 10^{3} - 10^{5}$ yr; ][]{1998PASP..110..493O,1999ASPC..193...79D} radio source where the radio structure does not extend beyond the optical scale of the galaxy. 

Studies of CSS objects have shown that most of them possess high velocity dispersion (FWHM $\sim 1000$ km s$^{-1}$) components, similar to 3C\,99 and 4C\,31.38 \citep[e.g.][]{2009MNRAS.400..589H, 2013ApJ...772..138S}. Most of the average EELR FWHMs found in this study are $< 600$  km s$^{-1}$, and none are $> 1000$  km s$^{-1}$. The higher velocity dispersion in the younger radio sources could be a result of shocking from either the radio jets themselves, or the accompanying high pressure bubble. In contrast, for the objects with more extended radio structures, while such interactions can occur outside of our field of view, none are detectable close to the nucleus. An alternative explanation is that, because of the more compact nature of EELRs around CSS objects, we are seeing a superposition of bulk motions over a large range of directions rather than a true velocity dispersion. 

\subsection{Velocity Field}

Previous studies of a few luminous EELRs such as \citet{2009ApJ...690..953F} have shown that while the EELR clouds show varying degrees of local organization, the whole velocity field cannot be described by one global dynamical model. In our sample, a number of velocity fields are consistent with large scale rotational motion, while others appear more complicated. The velocity structure of the EELRs around radio loud sources can be shaped by a few different events/mechanisms: (1) The radio jet depositing energy to the surrounding gas and causing outflow, (2) the disturbance from a recent merger, and (3) the host galaxy's own gas kinematics. The first event accounts for gas moving radially outward from the nucleus, and the two later ones account for non-outflow gas motions. 

In case (1) we may expect kinematics similar to scaled up narrow line regions (NLRs) affected by radio jets, which show kinematics that can be well described by a biconical outflow model \citep[e.g.,][]{2006AJ....132..620D}. However, our fitting result shows that only one velocity field can be fitted by a pure biconical model, indicating that other factors may be at work. 

Case (2) stems from the observational evidence which indicates that the host galaxies of the radio loud sources have likely been through a recent gas poor elliptical + gas rich disk merger, as discussed in \S\ \ref{discussion:detrate}. The rotational motion can come from the merging gas rich disk, and more chaotic motions can be attributed to disturbance from the merger. In the early stages after a merger, some gas is expected to lose its angular momentum and funnel into the center, but there will remain a significant fraction of gas in the tidal tail which may retain much of the initial rotation. If the radio source initiates the feedback during this time, the outflowing wind may sweep up the rotating gas in the outskirts without major disruptions to the rotating pattern. 

Another possibility is that the EELRs are distributions of interstellar gas simply illuminated by the AGN instead of outflow gas. Indeed, the gas velocities that are within $\sim 350$ km s$^{-1}$ do not require an outflow scenario to explain. If the EELRs are gas clouds bounded by gravity, both case (2) and (3) can apply. However, given the velocity offsets and velocity dispersions, the tails beyond the FWHM in the emission-line profiles can reach up to $> 600$ km s$^{-1}$ relative to the systemic velocity for most objects ($> 1000$ km s$^{-1}$ for several more extreme cases), which almost certainly exceed the escape velocities of the host galaxies. The escape velocity of a galaxy with $M = 10^{11.5} M_{\odot}$ at 5 kpc is around 585 km s$^{-1}$. Even though only a small fraction of gas may reach such high velocities, this suggests that non-gravitational forces are significant. It is also important to note that we are only observing the radial component of the velocities. The intrinsic gas velocities are likely higher and closer to the escape velocities of the host galaxies. 

Alternatively, the EELR can consist of both outflow and non-outflow gas. For example, the EELR of 3C\,249.1, studied in \citet{2006ApJ...650...80F} has one velocity component which shows rotational pattern and stays within $\sim 200$ km s$^{-1}$ of the systemic velocity, and two other velocity components with higher velocities. The first component can be from gas within the host galaxy and the other two may be outflowing gas. Similarly, the EELRs that appear to be rotating and/or have less extreme kinematics may be ionized gas within the host galaxy, while the jet-driven outflow contributes to more complicated patterns and higher velocities.

\subsection{Mass Outflow Rate}

The typical velocity of the EELR gas is around a few hundred km/s. The power of the outflow is $\dot{E} = \frac{1}{2} \dot{M} v^2$. Given an average mass outflow rate of 30 M$_\odot$/yr, and assuming an outflow velocity of 200 km s$^{-1}$, the power of the outflow is roughly $1.9 \times 10^{41}$ ergs s$^{-1}$. However, this value is likely based on an underestimated velocity. Considering the velocity dispersions and projection effects, the intrinsic average velocity of the gas should be higher. If we take the fitted model velocities from Table \ref{fitparam}, given that many objects have gas velocities $\sim 500$ km s$^{-1}$, the average outflow power can increase by almost an order of magnitude. The outflow speed is also expected to be higher at the earlier stages of the outflow. In these more evolved radio sources, the outflow rate have likely slowed down through mass entrainment and work against gravity.  The power of the outflow at earlier stages may be $1 - 2$ orders of magnitude higher than the power at later stages. 

The mass outflow rates are comparable to the broad-absorption-line (BAL) quasars, which have mass outflow rates of $10 - 100$ M$_\odot$/yr \citep[][and references therein]{2013MNRAS.436.3286A}. 

EELRs have also been found around radio quiet quasars with very high [\ion{O}{3}] luminosities ($L_{[O III]} > 10^{42.8}$. The mass and mass outflow rates of the EELRs in our sample are comparable to those calculated for the EELRs around radio quiet quasars in \citet{2013MNRAS.436.2576L} (Note: we are comparing our result to the values before extinction corrections in section 6.1 of \citet{2013MNRAS.436.2576L}, since we do not have the emission lines needed to estimate extinctions for our objects).

\section{Conclusions and Future Work}

We have carried out the first study of EELRs in a matched sample of quasars and radio galaxies, analyzing IFU and long-slit data to study the distribution and velocity field of the ionized gas. We found no significant difference between the detection rate of EELRs around the two populations. The distribution of velocity dispersions and velocity structures is also similar for the EELRs around the quasars and the radio galaxies. This is consistent with widespread extended emission-line clouds detectable from all viewing angles and supports the AGN unified scheme.

The projected morphology of EELRs spans a range from a narrow elongation to a clumpy wide-angle distribution. Regardless of the apparent opening angle, there is no preferential alignment with the radio axes.  This shows that, to first order, the 3-D morphologies of the EELRs are closer to being spherically symmetric than confined in narrow ($< 45$\arcdeg) cones around the radio axes. However, the quasars do appear more likely to have EELRs that extend out in all directions, suggesting a wide-angle ($> 45$\arcdeg) biconical outflow where the cones are pointing closer to our line of sight. 

The EELR velocity field structure ranges from simple large scale rotations to complicated patterns that cannot be explained by simple dynamical models. A pure biconical outflow model seems to apply to only one case. The rotating disk, on the other hand, can provide good fits for $\sim 1/4$ of the EELRs found in our sample. Our result agrees with previous observations which show that the radio axes do not appear to correlate with the velocity structure. The rotation patterns are possibly inherited from a gas rich member of a merger which triggered the AGN. 

Combining the results from this study and other observations of emission-line regions around radio loud sources, the evolutionary trends appear to be the following: As the radio sources become older and more extended (1) the size of the EELRs becomes larger \cite[e.g.,][]{2013ApJ...772..138S}, (2) the overall velocity dispersion of the EELRs decreases \citep[e.g.,][]{2009MNRAS.400..589H, 2013ApJ...772..138S}, and (3) the morphological and kinematic alignment between the EELRs and the radio structure becomes less pronounced \cite[e.g.,][]{2008ApJS..175..423P,2013ApJ...772..138S}. While there is little evidence of direct interaction between the EELRs and the radio jets, the strong correlation between luminous EELRs and FR II sources, as well as the evolutionary trends suggest that the radio jets do play an important role in shaping the EELRs. 

Feedback is widely accepted to be a crucial element in the scheme of galaxy evolution, and outflows are generally considered observational evidence of feedback in progress. Current evidence supports the outflow origin for the EELRs (See discussion in e.g., \citet[][]{2007ApJ...666..794F}). In our sample, some EELRs gas show low velocities with respect to the systemic velocity, which may be bounded by gravity. However, the observed gas velocities are only the radial components, and the intrinsic velocities are likely higher. Also, given the velocity dispersions, the gas in some objects almost certainly reach above the escape velocities. Therefore our results suggest that EELRs do represent one form of the feedback process. 

We estimated an average mass of $\sim 3 \times 10^8$ M$\odot$ for the EELRs and a mass outflow rate of $\sim 30$ M$_\odot$/yr. This is comparable to other ionized outflows such as those associated with BALs and radio quiet objects, suggesting that the radio jet driven outflows have as much impact on their host galaxies as the other forms of ionized outflows. 

Higher spectral/spatial resolution and coverage IFU observations have only been done for a few EELRs. Obtaining such observations of the EELRs identified in this study will help distinguish the different velocity components often observed in luminous EELRs, allow better determination of the fraction of high velocity gas, improve our velocity model fits, and provide information such as H$\beta$ luminosities that allow more accurate mass calculation. These information will further constrain the potential influence the EELRs can have on their host galaxies.

\section{Acknowledgement}
We thank the anonymous referee for the useful comments that helped us improve this paper. This research has been partially supported by NSF grant AST-0807900. The authors wish to recognize and acknowledge the very significant cultural role and reverence that the summit of Mauna Kea has always had within the indigenous Hawaiian community.  We are most fortunate to have the opportunity to conduct observations from this mountain.

\begin{figure*}[t]
\centering
\includegraphics[ width=6.in]{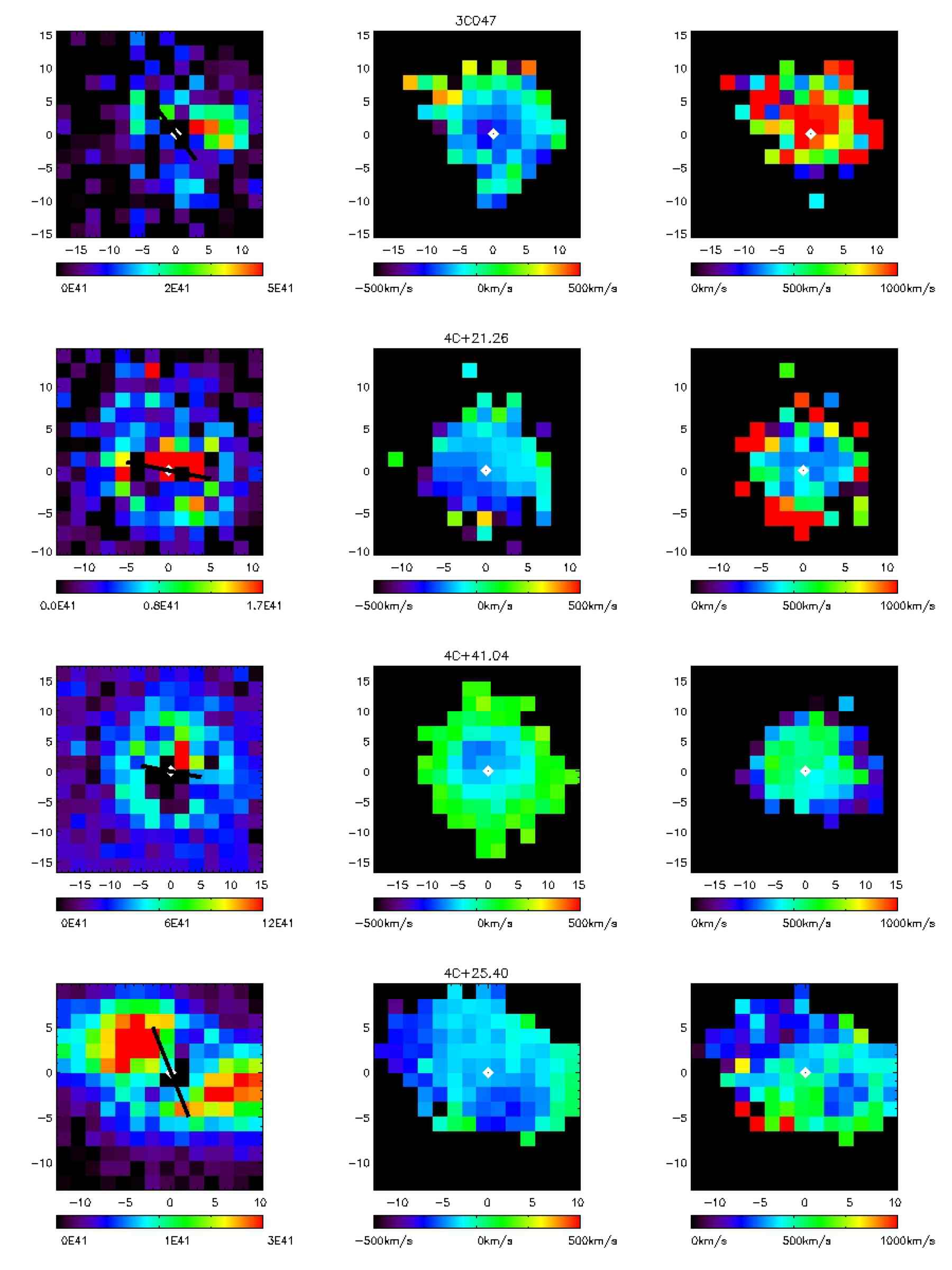}
\caption{EELR flux distribution in erg s$^{-1}$ arcsec$^{-2}$ (left), velocity (middle) and velocity dispersion (FWHM) (right) maps. The black line in the left column indicated the direction of the radio axis. The radio sources vary over two orders of magnitude in angular sizes, with most extending well beyond the SNIFS field of view. The velocities are calculated relative to the systemic velocities given in NED, which may be offset from the true systemic velocities by a few hundred km s$^{-1}$.}
\figurenum{4}
\label{eelrmaps0}
\end{figure*}

\begin{figure*}[t]
\centering
\includegraphics[width=6.in]{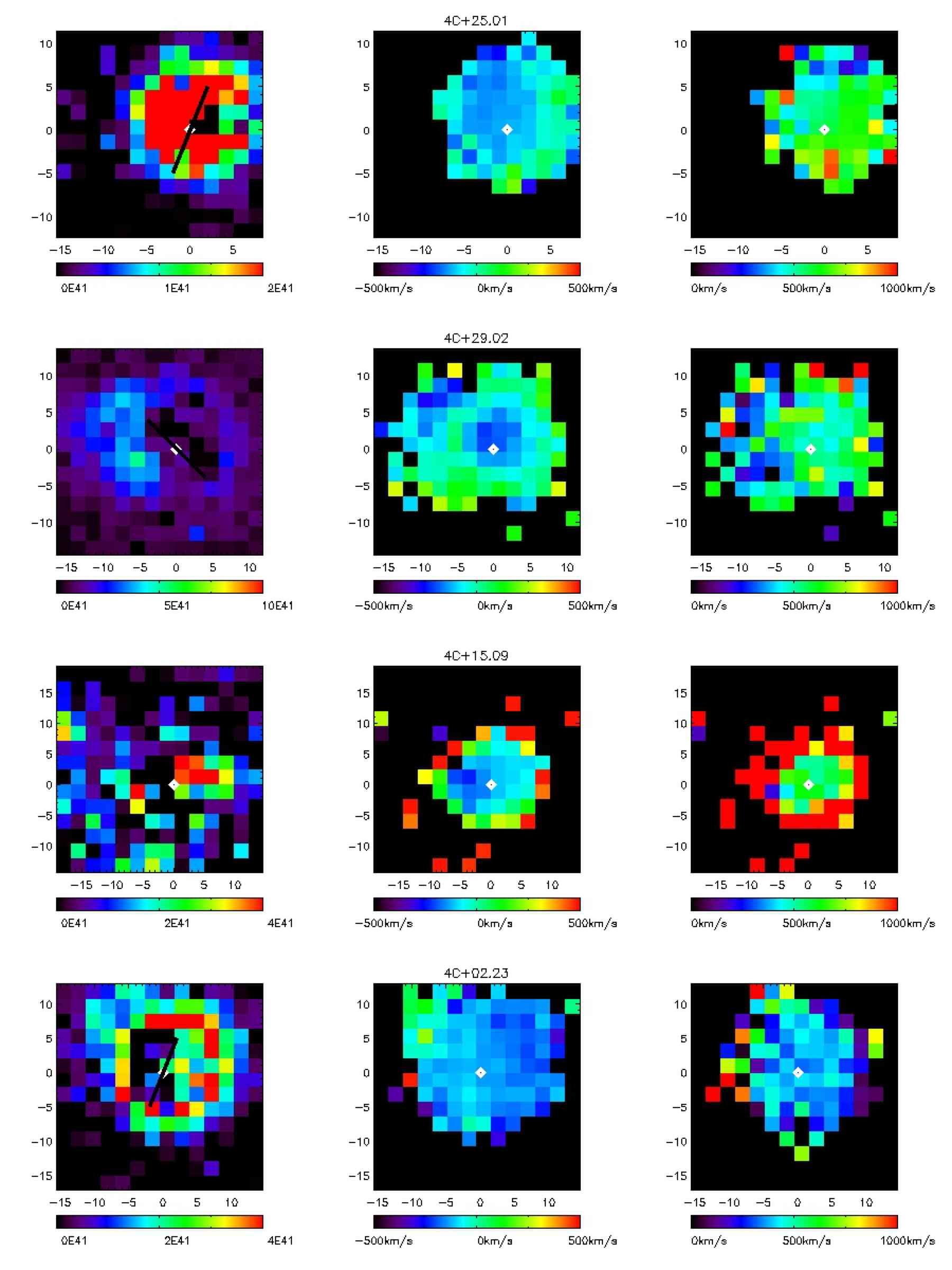}
\caption{(continued)}
\figurenum{4}
\label{eelrmaps1}
\end{figure*}

\begin{figure*}[t]
\centering
\includegraphics[width=6.in]{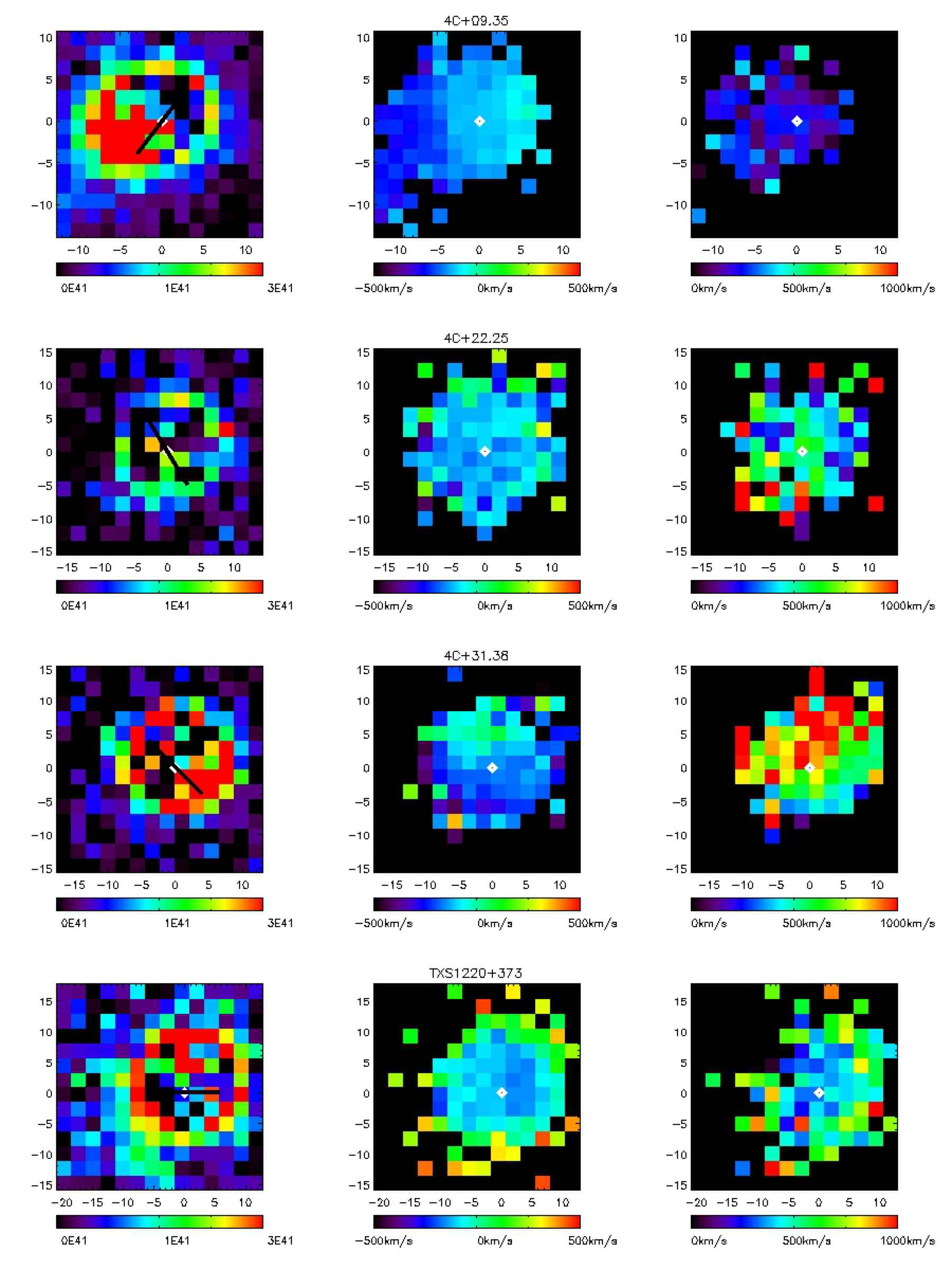}
\caption{(continued)}
\figurenum{4}
\label{eelrmaps2}
\end{figure*}

\begin{figure*}[t]
\centering
\includegraphics[width=6.in]{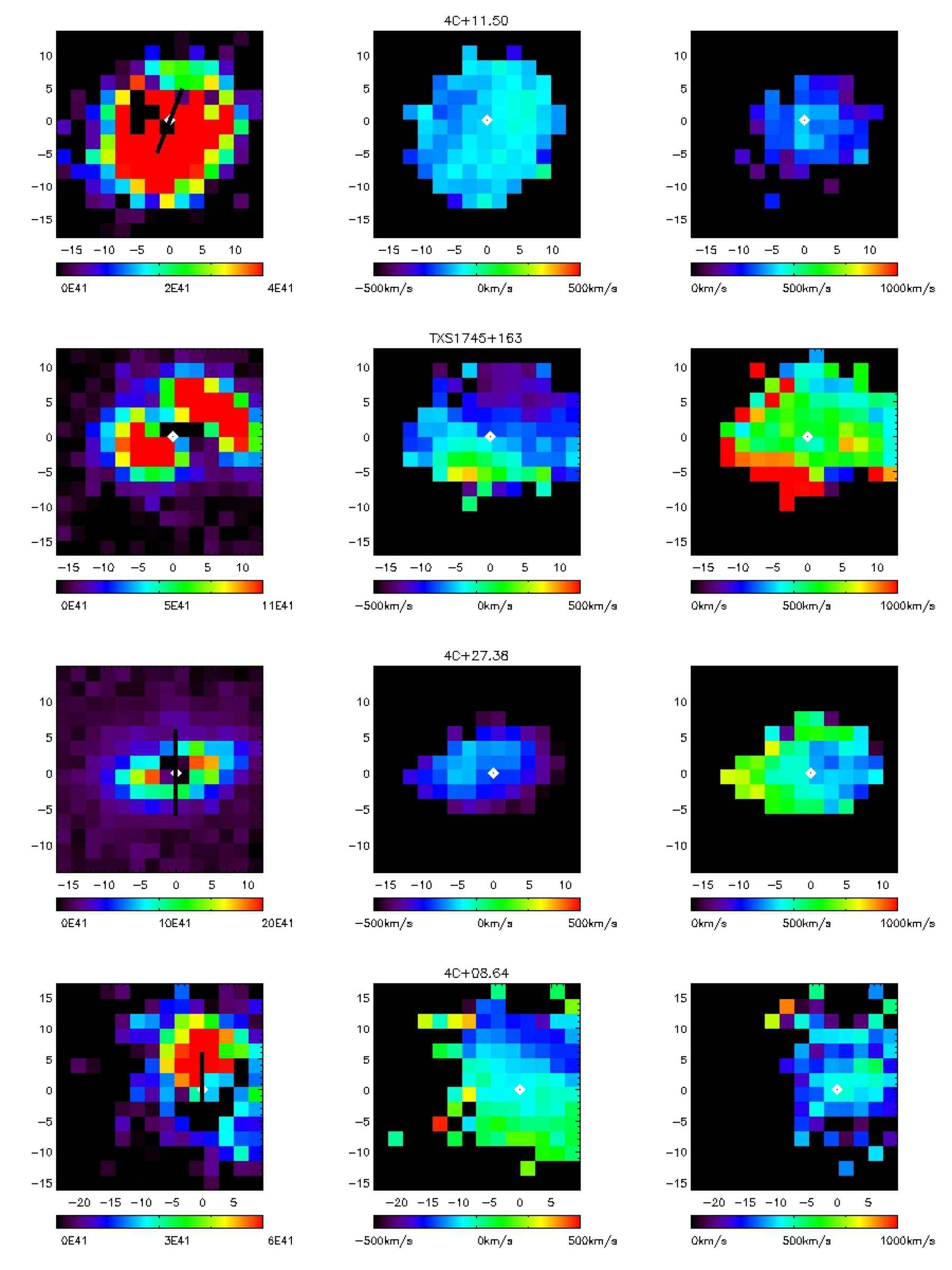}
\caption{(continued)}
\figurenum{4}
\label{eelrmaps3}
\end{figure*}

\begin{figure*}[t]
\centering
\includegraphics[width=6.in]{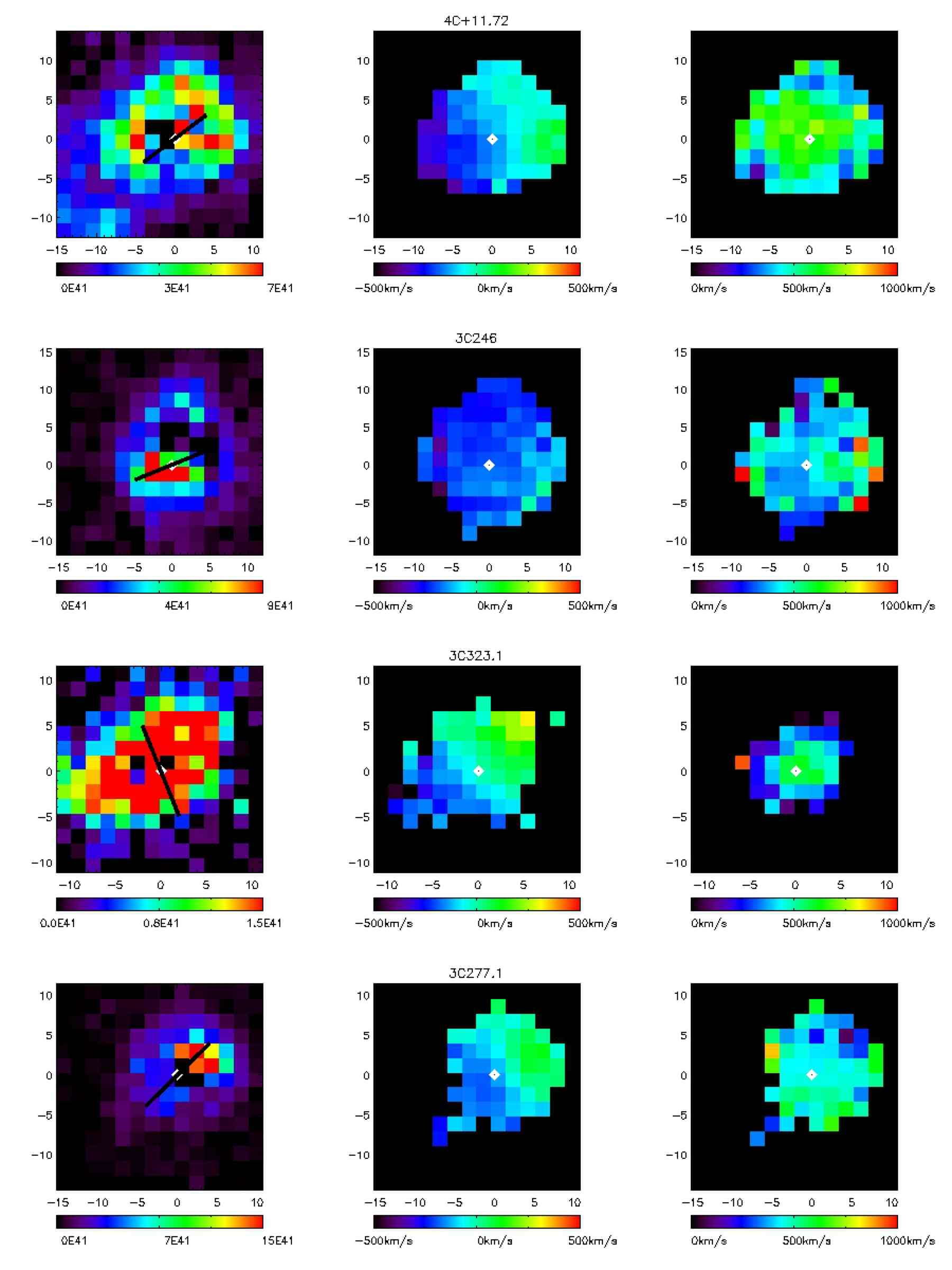}
\caption{(continued)}
\figurenum{4}
\label{eelrmaps4}
\end{figure*}

\begin{figure*}[t]
\centering
\includegraphics[width=6.in]{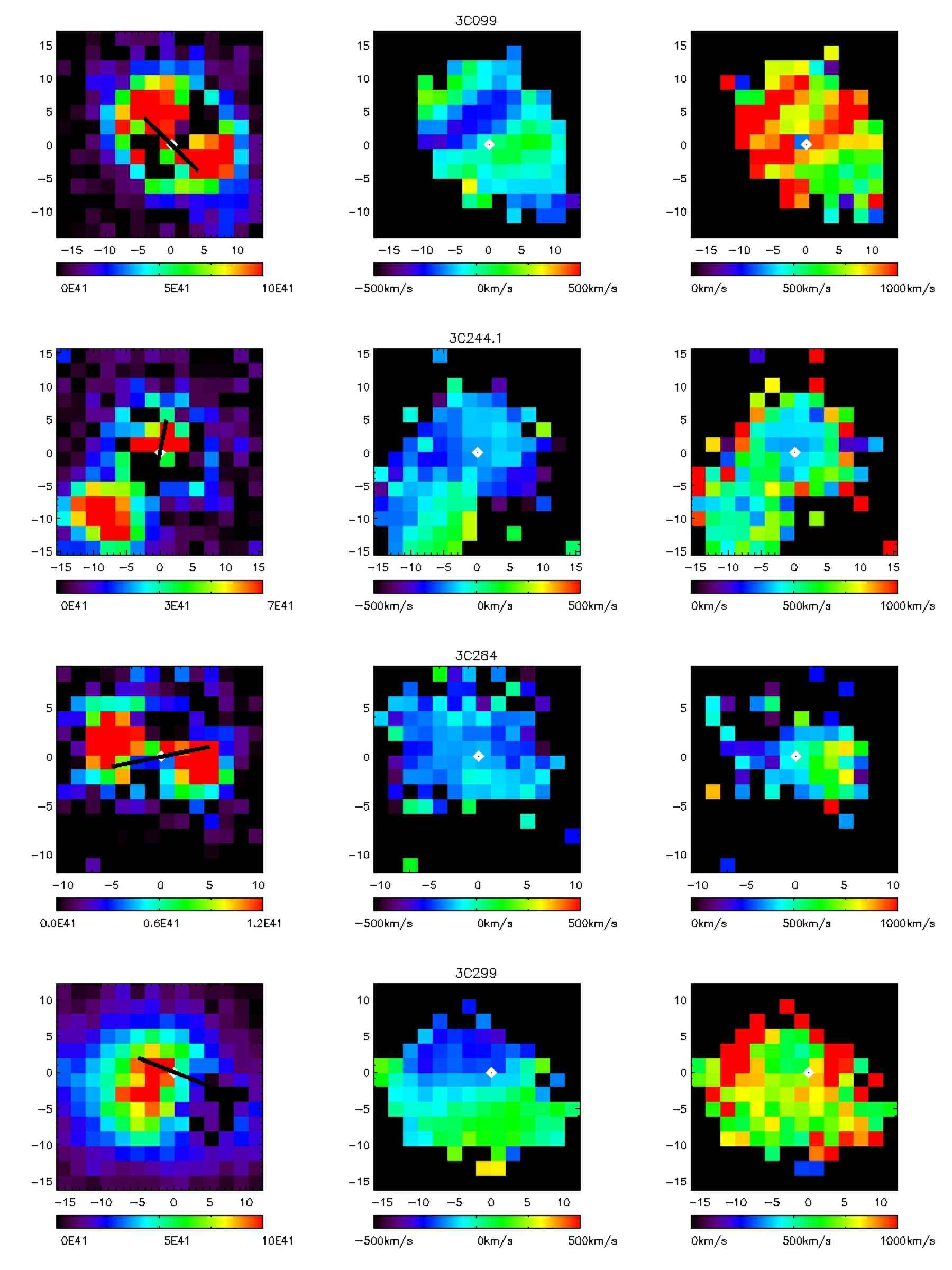}
\caption{(continued)}
\figurenum{4}
\label{eelrmaps5}
\end{figure*}

\begin{figure*}[t]
\centering
\includegraphics[width=6.in]{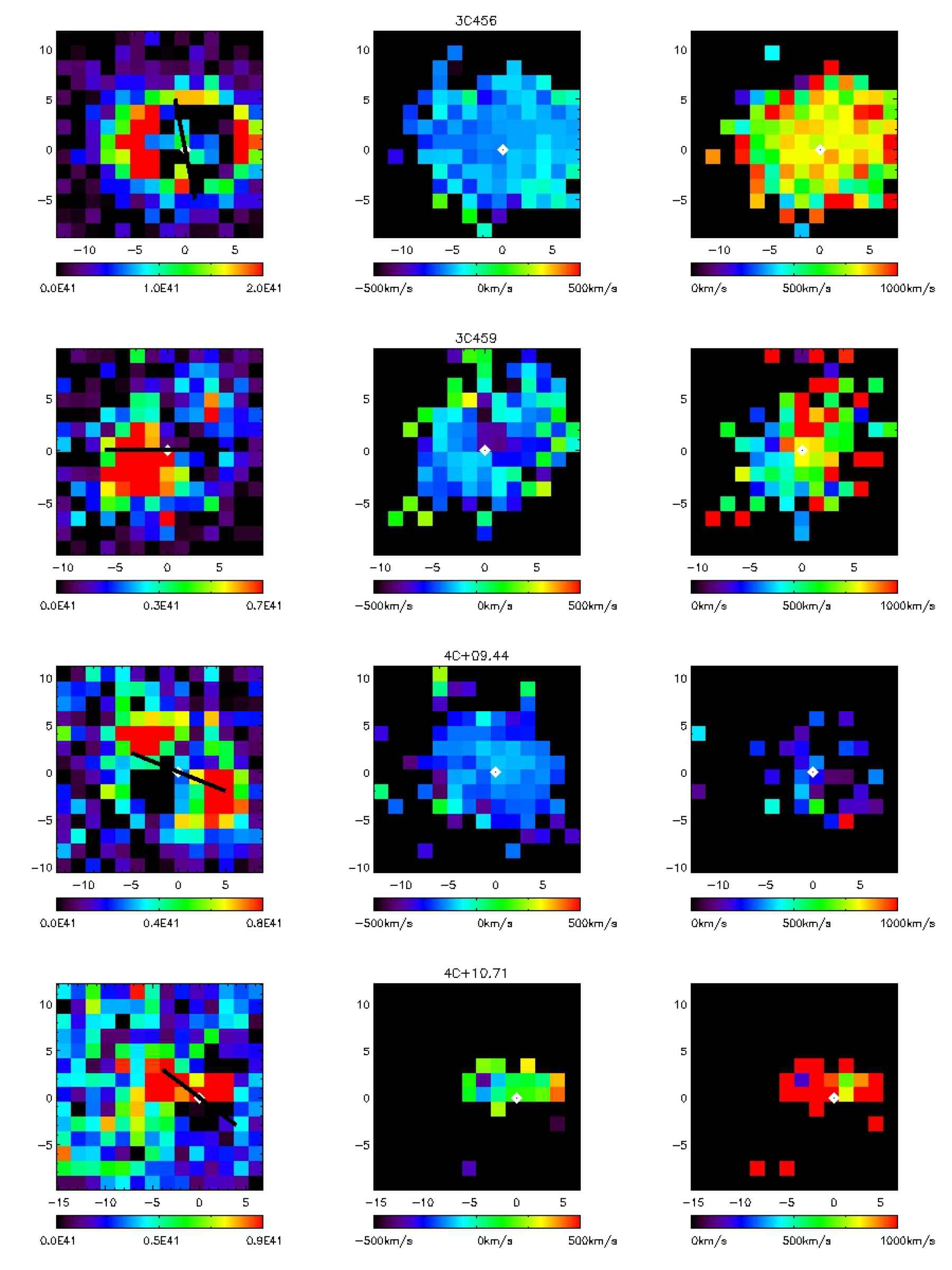}
\caption{(continued)}
\figurenum{4}
\label{eelrmaps6}
\end{figure*}


\begin{figure*}[t]
\centering
\includegraphics[width=6.in]{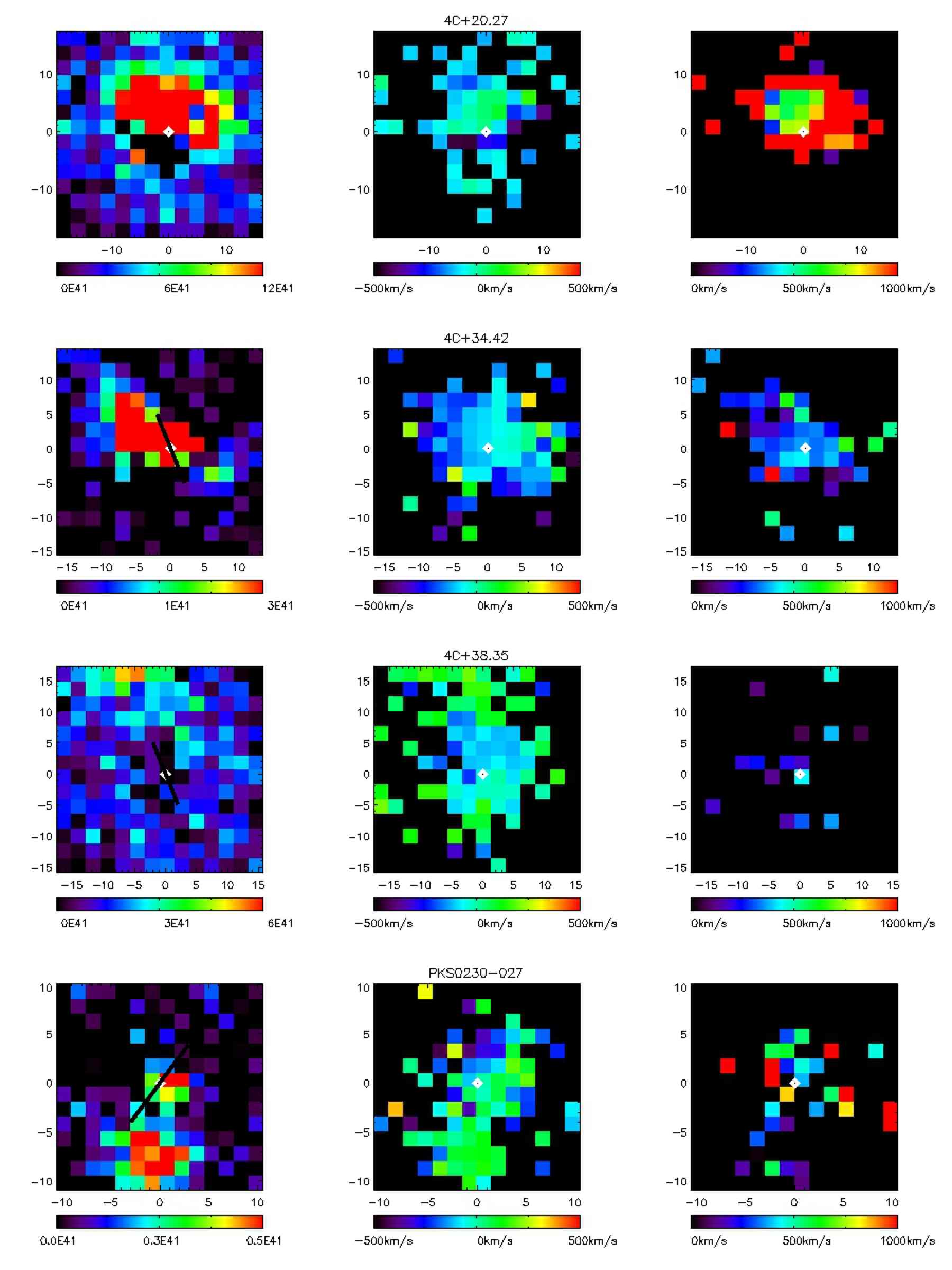}
\caption{(continued)}
\label{eelrmaps7}
\end{figure*}

\begin{figure*}[t]
\centering
\includegraphics[width=6.in]{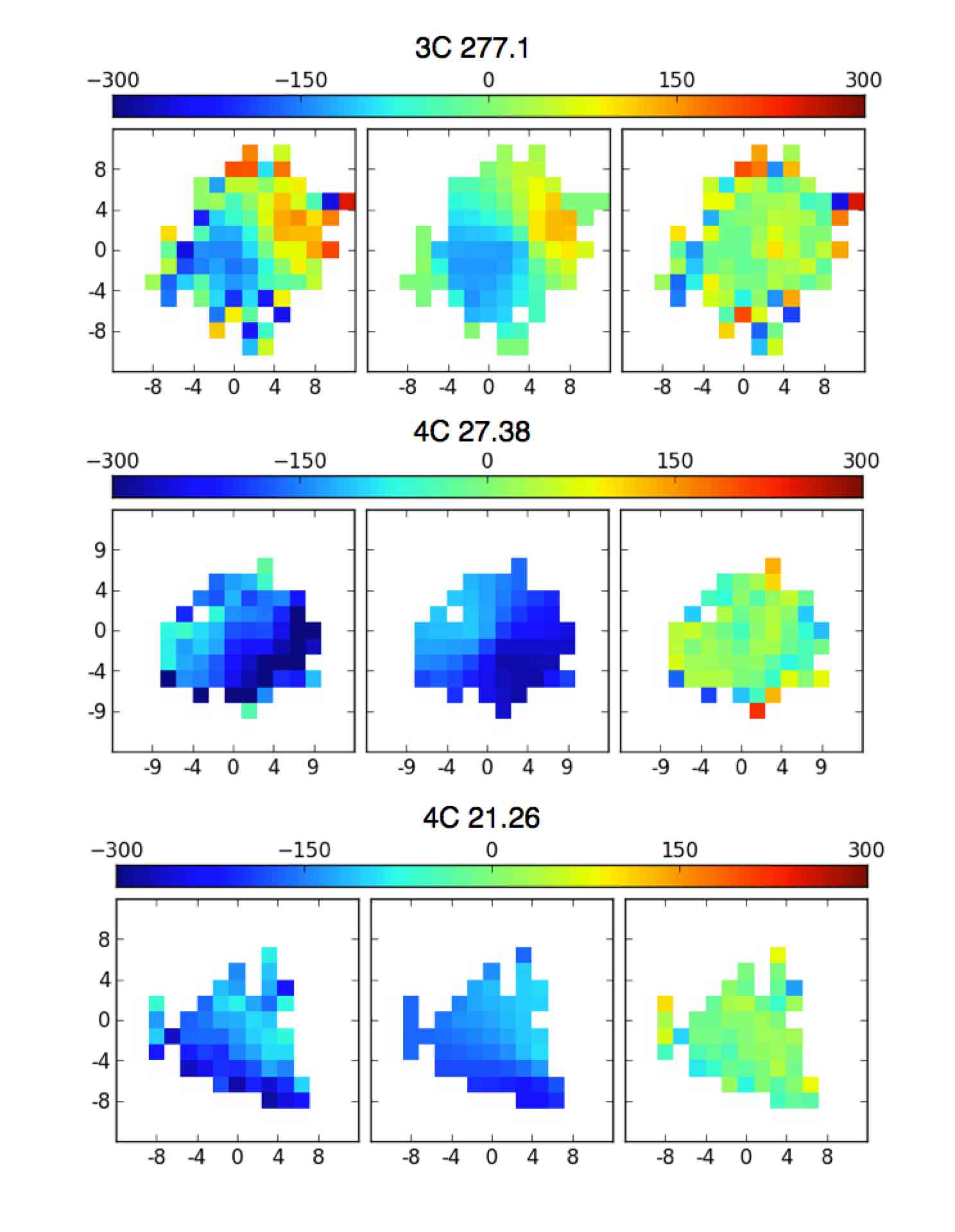}
\caption{Velocity map model fits for three objects. Approximate distances from center are marked in units of kpc on the x and y axes of each panel. Left column: Data. Middle column: Model. Right column Residual. The labels on the color bars are in units of km /s. }
\figurenum{5}
\label{modelfit1}
\end{figure*}

\begin{figure*}[t]
\centering
\includegraphics[width=6.in]{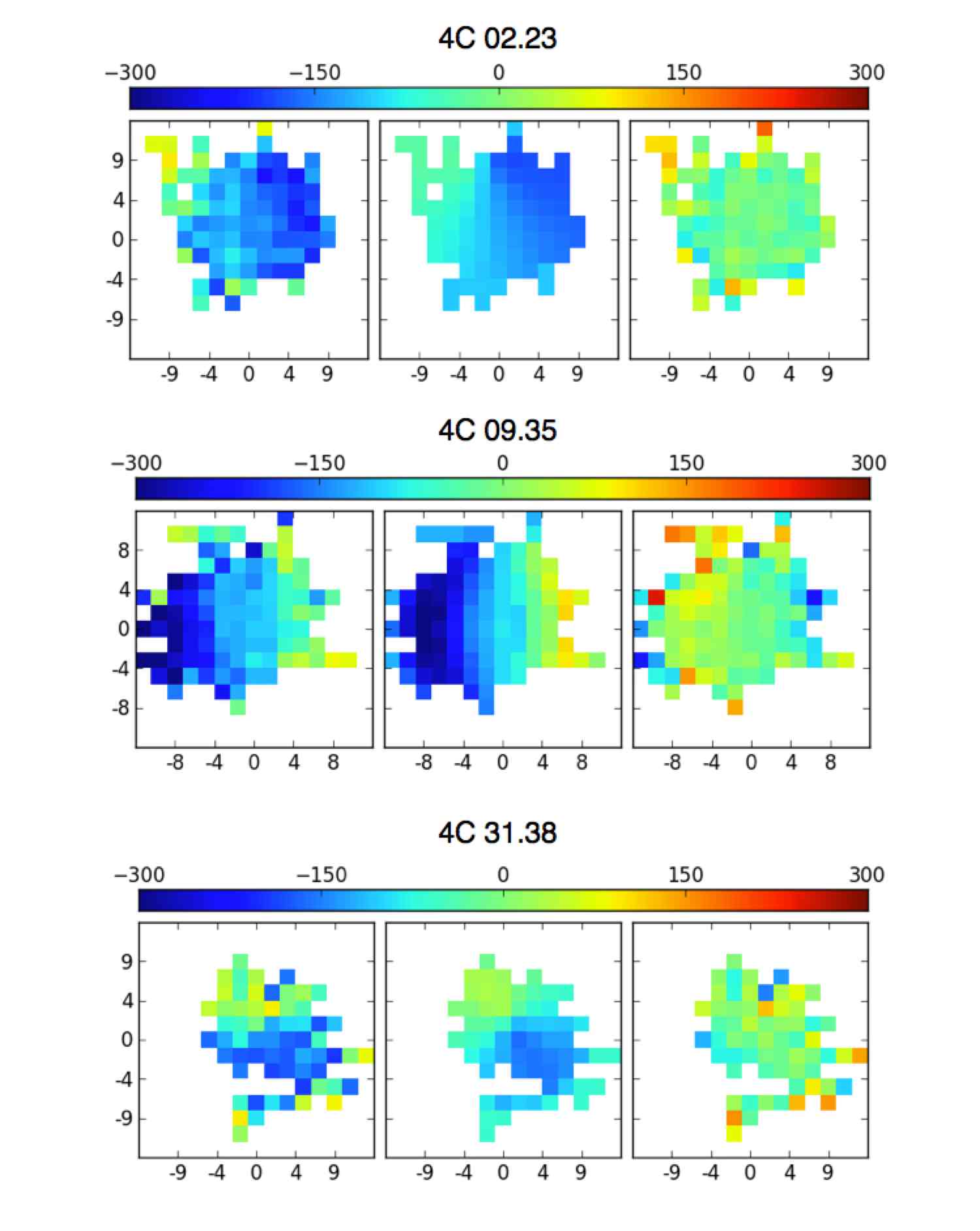}
\caption{(Continued)}
\figurenum{5}
\label{modelfit2}
\end{figure*}

\begin{figure*}[t]
\centering
\includegraphics[width=6.in]{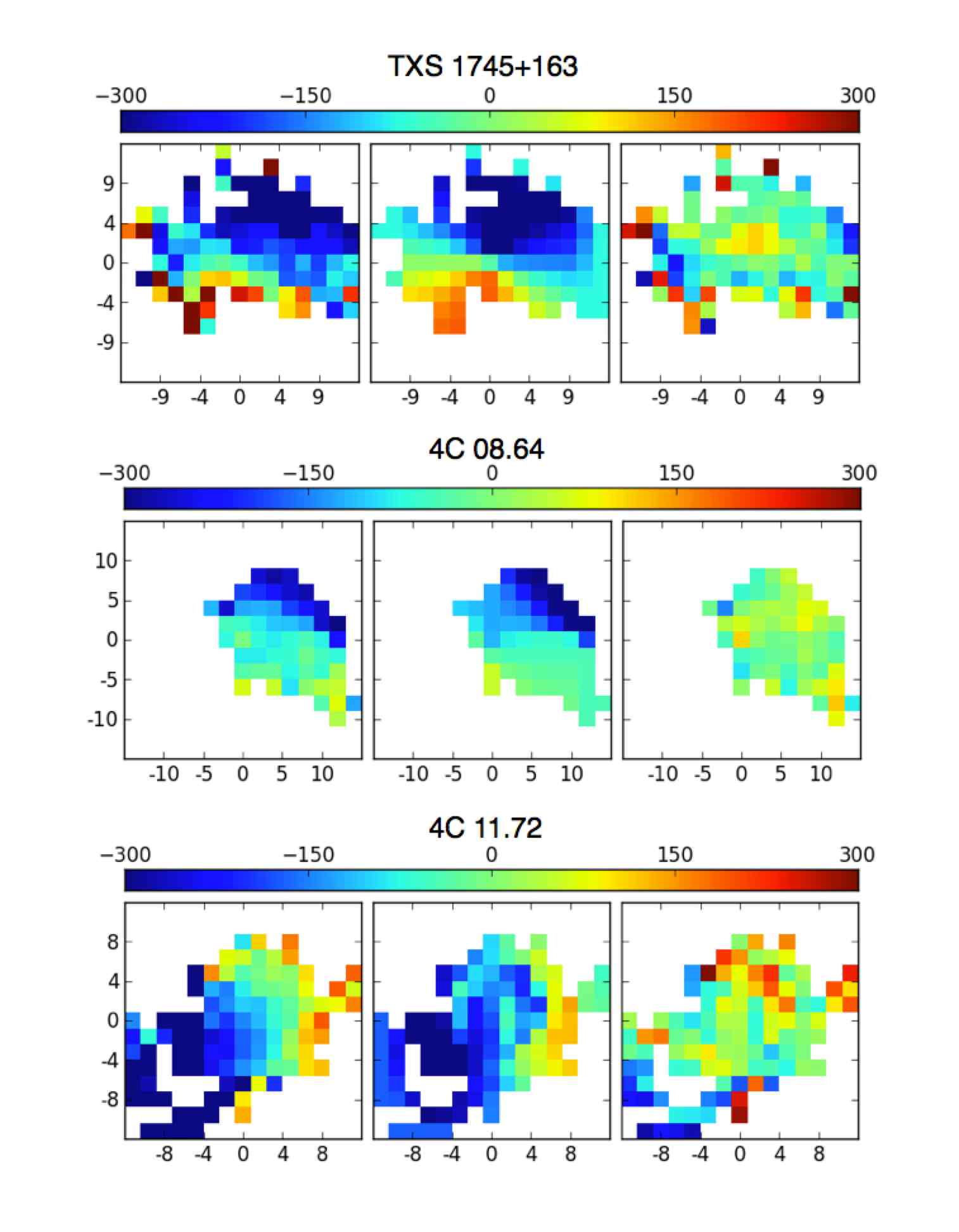}
\caption{(Continued)}
\figurenum{5}
\label{modelfit3}
\end{figure*}

\begin{figure*}[t]
\centering
\includegraphics[width=6.in]{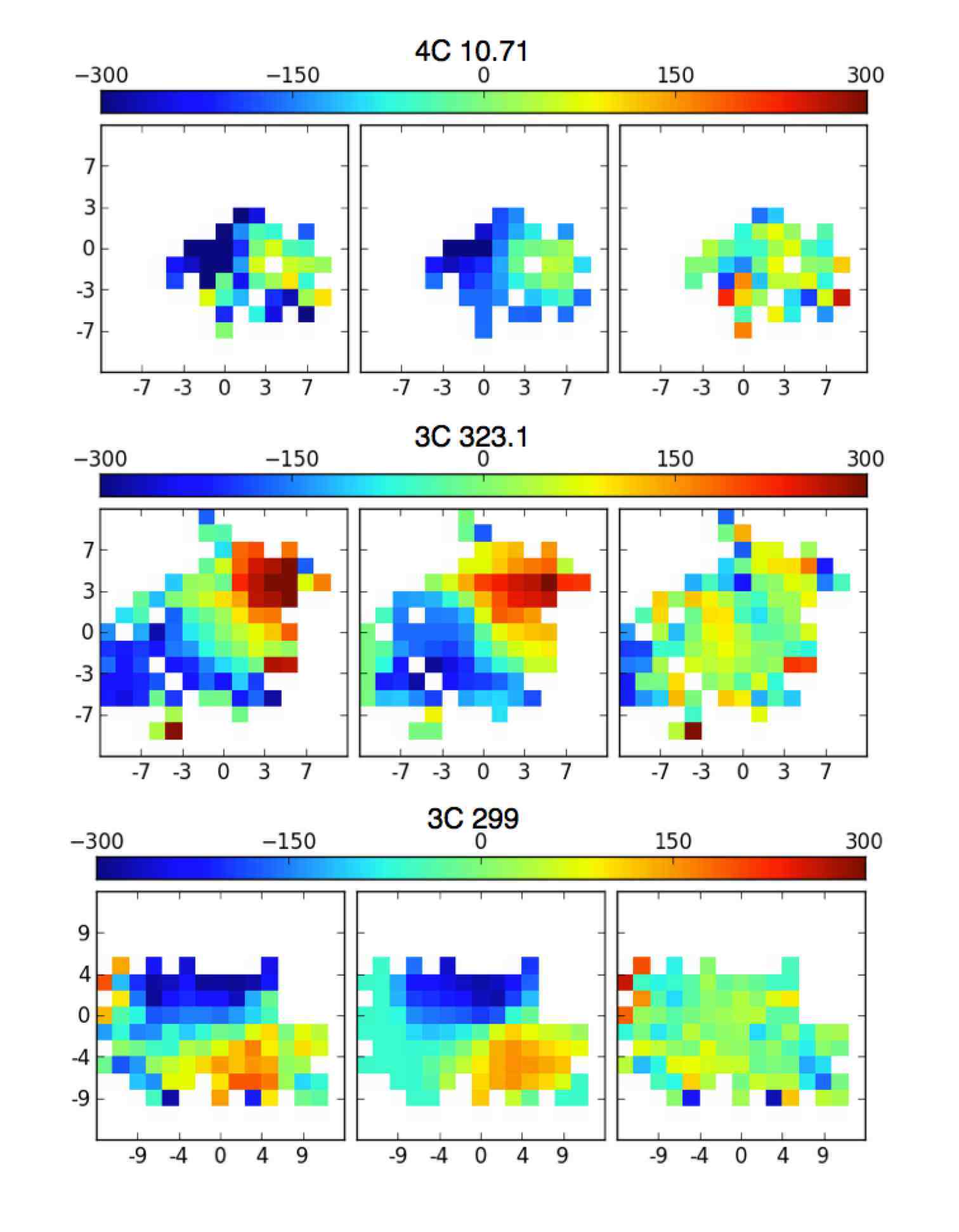}
\caption{(Continued)}
\figurenum{5}
\label{modelfit4}
\end{figure*}

\pagebreak

\bibliographystyle{apj}
\bibliography{/Users/hsshih/Dropbox/bib.bib}

\end{document}